\documentclass[aps,prb,twocolumn,floats,fleqn,eqsecnum,floatfix]{revtex4}

\usepackage{amsmath}
\usepackage{graphicx}

\begin{document}

\title{The Hubbard model with intersite interaction \\ within the Composite Operator Method}
\author{Adolfo Avella}
\email[E-mail: ]{avella@sa.infn.it}
\author{Ferdinando Mancini}
\email[E-mail: ]{mancini@sa.infn.it}
\affiliation{Dipartimento di Fisica ``E.R. Caianiello'' - Unit\`a di Ricerca INFM di Salerno\\
University\`a deli Studs di Salerno, I-84081 Baroness (SA), Italy}
\date{\today}

\begin{abstract}
We study the one- and two- dimensional extended Hubbard model by
means of the Composite Operator Method within the 2-pole
approximation. The fermionic propagator is computed fully
self-consistently as a function of temperature, filling and
Coulomb interactions. The behaviors of the chemical potential
(global indicator) and of the double occupancy and
nearest-neighbor density-density correlator (local indicators) are
analyzed in detail as primary sources of information regarding the
instability of the paramagnetic (metal and insulator) phase
towards charge ordering driven by the intersite Coulomb
interaction. Very rich phase diagrams (multiple first and second
order phase transitions, critical points, reentrant behavior) have
been found and discussed with respect to both metal-insulator and
charge ordering transitions: the connections with the experimental
findings relative to some manganese compounds are analyzed.
Moreover, the possibility of improving the capability of
describing cuprates with respect to the simple Hubbard model is
discussed through the analysis of the Fermi surface and density of
states features. We also report about the specific heat behavior
in presence of the intersite interaction and the appearance of
crossing points.
\end{abstract}

\pacs{???}

\maketitle

\section{Introduction}
\label{Intro}

Many authors have emphasized the importance of considering
non-local Coulomb interactions in describing doped systems like
cuprate superconductors or fullerides
\cite{Varma:87,Littlewood:89,Varma:95,Janner:95,vandenBrink:95}.
The simplest Hamiltonian satisfying these requirements is the
extended Hubbard model, where a nearest-neighbor Coulomb
interaction term is added to the original Hubbard
Hamiltonian\cite{Hubbard}. The inclusion of non-local Coulomb
interactions substantially modifies the electronic properties of
the model. For instance, the charge transfer excitons, which can
only be detected by optical spectroscopy at half filling, attain
some charge in doped systems and become visible in direct and
inverse photoelectron spectroscopies \cite{vandenBrink:97}. Other
studies, using an effective extended Hubbard model, support the
appearance, upon doping, of states evenly distributed inside the
gap \cite{Simon:97}. This suggests that the general features of
the cuprates can be well described by using this effective Hubbard
Hamiltonian, which has also been used to mimic some of their
experimental features in the superconducting state by means of a
BCS treatment \cite{Ferrer:98}.

The Hubbard model with intersite Coulomb interaction is also one
of the simplest models capable to describe charge ordering (CO) in
interacting electron systems. Already in 1938 Wigner proposed
\cite{Wigner:38} that a low density interacting electron gas
crystallizes in a lattice in order to minimize the Coulomb
repulsion. At higher densities crystallization is possible if the
kinetics energy is reduced by spin or phonon interactions
\cite{Fulde:97}. Charge ordering has been experimentally observed
in a variety of systems: $GaAs/AlGaAs$ heterostructures
\cite{Andrei:88}, rare-earth pnictides like $Yb_4As_3$
\cite{Ochiai:90}, colossal magnetoresistance compounds
\cite{Chen:96}, unconventional spin-Peierls materials $\alpha
-NaV_2O_5$ \cite{Ohama:99}, cuprates \cite{Vojta:00}, manganites
\cite{Salamon:01}, magnetite \cite{Park:98}, vanadium oxides
\cite{Ueda:01}, Bechgaard salts \cite{Chow:00}.

Among the many analytical methods used to study the extended
Hubbard model we recall: Hartree-Fock approximation \cite{Seo:98},
perturbation theory \cite{vanDongen:94a}, dynamical mean field
theory \cite{Pietig:99}, slave boson approach
\cite{McKenzie:01,Merino:01}, coherent potential approximation
\cite{Hoang:02}. Numerical studies by means of Quantum Monte Carlo
\cite{Hirsch:84a}, Lanczos technique \cite{Hellberg:01} and exact
diagonalization \cite{Calandra:02} have also to be recalled.

In this manuscript, we analyze the extended Hubbard model by means
of the Composite Operator Method (COM) within the 2-pole
approximation (see Refs.~\onlinecite{Mancini:00,Mancini:04} and
references therein).

The COM is based on two main concepts: (i) the excitations present
in interacting systems are far from being the original electrons
and need to be described by asymptotic fields in the form of
composite operators; (ii) the use of composite operators requires
the enforcement of the non-canonical algebra they obey in order to
properly fix the representation where their propagators are
realized. This latter task is effectively undertaken by computing
the values of the unknown correlators appearing in the
calculations by means of Algebra Constraints \cite{Mancini:00}.

The detailed analysis of the instabilities of the homogenous
paramagnetic phase of the extended Hubbard model towards charge
ordered inhomogeneous phases is the main task undertaken with this
manuscript. With respect to this, we have analyzed the rank of the
transitions and their relations with the metal-insulator one.
Actually, this study is the only relevant at room temperatures
and/or in presence of frustration as any spin ordered phase (e.g.,
the antiferromagnetic phase) would be inaccessible in such
situations. It is worth noticing that the very rich phase diagrams
(multiple first and second order phase transitions, critical
points, reentrant behavior) that have been found can be put in
connection with the experimental findings relative to some
manganese compounds \cite{Tomioka:97,Chatterji:00,Dho:01}. We have
also discussed in detail the possibility of improving the
capability of describing cuprates with respect to the simple
Hubbard model and the appearance of crossing points in the
specific heat in presence of the intersite interaction.

In the next section we present the model, the basis and the
solution according to the COM. In the subsequent sections, we
comment our results for the chemical potential, the phase diagram,
the double occupancy, the nearest-neighbor density-density
correlator, the internal and kinetic energies, the Fermi surface,
the density of states and the specific heat. The types of charge
ordered phases found according to the sign of the intersite
potential are described, the rank of the transitions is evidenced
through the analysis of the discontinuities in the chemical
potential, double occupancy, nearest-neighbor density-density
correlator and kinetic and internal energies, quite complex phase
diagram (with metal to insulator and to charge ordered phase
transitions and reentrant behavior) are drawn and commented, the
behavior of single- (double occupancy) and two- (nearest-neighbor
charge) site correlators is studied in order to get information
about the actual charge distribution, the value of the filling at
which the nesting appear is determined as a function of intersite
potential, the specific heat behavior is studied in comparison to
that of the simple Hubbard model.

\section{Hamiltonian, field equations and solution}

We will study a generalized version of the Hubbard model
\cite{Hubbard} which includes the intersite Coulomb interaction
\cite{Hirsch:84,Emery:79,Emery:87,Zhang:88,Bosch:88,Zhang:89,Yan:93}.
Accordingly, the Hamiltonian under analysis reads as
\begin{align}\label{Eq.5.1.1}
\begin{split}
H&=\sum_{\bf i} [-\mu c^\dagger (i)c(i)-2dtc^\dagger (i)c^\alpha
(i)]\\
&+\sum_{\bf i} [Un_\uparrow (i)n_\downarrow (i)+dVn(i)n^\alpha
(i)]
\end{split}
\end{align}
where $\mu$ is the chemical potential,
$c^\dagger(i)=(c_\uparrow^\dagger(i) \, c_\downarrow^\dagger(i))$
is the creation electronic operator in spinorial notation,
$i=(\mathbf{i}, t)$, $\mathbf{i}$ is one lattice vector of the
$d$-dimensional square lattice, $t$, as usually done in the
related literature, is both the time variable and the hopping
integral, the context will clarify the use, $U$ is the onsite
Coulomb interaction, $V$ is the intersite interaction,
$n(i)=n_\uparrow(i)+n_\downarrow(i)$, $n_\sigma(i)$ is the number
operator for electrons of spin $\sigma$. Hereafter, $t$ will be
used as reference unit for all energies. We have used the notation
\begin{equation}
\phi^\alpha (\mathbf{i}, t)= \sum_{\bf j} \alpha_\mathbf{ij} \phi
(\mathbf{j}, t)
\end{equation}
where $\phi$ can be any operator and $\alpha_\mathbf{ij}$ is the
projector on the first $2d$ neighbor sites on the lattice. We have
$\alpha(\mathbf{k})=\mathcal{F}[\alpha_\mathbf{ij}]=1/d\sum_{n=1}^d
\cos(k_n)$, where $\mathcal{F}$ is the Fourier transform.

Within the Composite Operator Method\cite{Mancini:00}, once we
choose a $n$-component spinorial basis $\psi(i)$, the equations of
motion of this latter take the general form
\begin{equation}
\mathrm{i}\frac{\partial}{\partial t}\psi({\bf i},t)=\sum_{\bf
j}\epsilon({\bf i},{\bf j}) \psi({\bf j},t)+\delta j({\bf i},t)
\end{equation}
where $\epsilon({\bf k})=\mathcal{F}[\epsilon({\bf i},{\bf j})]$
is the $n \times n$ energy matrix describing the projected
dynamics. The energy matrix can be computed as $\epsilon({\bf
k})=m({\bf k})I^{-1}({\bf k})$ where $I({\bf
k})=\mathcal{F}\langle \{ \psi({\bf i},t), \psi^{\dagger}({\bf
j},t) \} \rangle$ is the normalization matrix of the basis and
$m({\bf k})=\mathcal{F}\langle \{
\mathrm{i}\frac{\partial}{\partial t}\psi({\bf i},t),
\psi^{\dagger}({\bf j},t)\} \rangle$. If we neglect $\delta j(i)$
we obtain a pole structure for the retarded thermal Green's
function $G({\bf k},\omega)=\mathcal{F}\langle \mathcal{R}[\psi
(i)\psi ^\dagger (j)] \rangle$ ($\mathcal{R}$ is the retarded
operator) that will obey the following equation of motion
\begin{equation}\label{GFF}
\omega G({\bf k},\omega)=I({\bf k})+\epsilon({\bf k})G({\bf
k},\omega)
\end{equation}
The solution of Eq.~(\ref{GFF}) is
\begin{equation}\label{2pole}
G(\omega, {\bf k})=\sum_{i=1}^n\frac{\sigma^{(i)}({\bf k})}{\omega
-E_{i}({\bf k})+\mathrm{i}\delta}.
\end{equation}
where $E_{i}({\bf k})$ are the eigenvalues of $\epsilon({\bf k})$
and the spectral weights $\sigma^{(i)}({\bf k})$ can be computed
as
\begin{equation} \label{Eq.B.19}
\sigma^{(i)}_{ab}({\bf k})=\sum_{c=1}^n \Omega_{ai}({\bf
k})\Omega_{ic}^{-1}({\bf k})I_{cb}({\bf k}) \quad a,b=1,\ldots,n
\end{equation}
where the matrix $\Omega ({\bf k})$ has the eigenvectors of
$\epsilon({\bf k})$ as columns\cite{Mancini:00}. In this
manuscript, we will use a 2-pole approximation within the COM. The
reader interested to more elaborate self-energy treatments (in
order to take into account $\delta j(i)$ we should compute the
higher order propagator $\langle \mathcal{R}[\delta j (i) \delta j
^\dagger (j)] \rangle$) should refer, for instance, to
Ref.~\onlinecite{Matsumoto:96,Matsumoto:97,Avella:03c}.

For the model under analysis in this manuscript, we choose, as
basic fields, the Hubbard operators, which faithfully describe the
Hubbard subbands as eigenoperators of the ionic model,
\begin{equation}\label{Eq.5.1.2}
\psi(i)= \begin{pmatrix} {\xi (i)}\cr {\eta (i)}
\end{pmatrix}
\end{equation}
where $\xi(i)=n(i)c(i)$ and $\eta(i)=c(i)-\xi(i)=[1-n(i)]c(i)$.
They satisfy the following equations of motion
\begin{multline}\label{Eq.5.1.3}
\mathrm{i}{\partial \over {\partial t}}\psi (i) \\
= \begin{pmatrix}
{-\mu \xi (i)-2d[tc^\alpha (i)+t\pi (i)-V\xi (i)n^\alpha (i)]}\\
{-(\mu -U)\eta (i)+2d[t\pi (i)+V\eta (i)n^\alpha (i)]}
\end{pmatrix}
\end{multline}
where
\begin{equation}\label{Eq.5.1.4}
\pi (i)={1 \over 2}\sigma ^\mu n_\mu (i)c^\alpha
(i)+c(i)c^{\alpha^\dagger }(i)c(i)
\end{equation}
$\sigma ^\mu=(-1,\vec{\sigma})$, $n_\mu(i)=(n(i),\vec{n}(i))$ is
the charge and spin number operator, $\vec{n}(i)=c^\dagger(i)
\vec{\sigma} c(i)$ and $\vec{\sigma}$ are the Pauli matrices.

After the choice we made for the basis, we can compute, in the
2-pole approximation within COM, the energy spectra $E_i({\bf k})$
and the spectral density functions $\sigma ^{(i)}({\bf k})$
according to the general procedure given above. The lengthy
expressions can be found in Appendix. It is worth noting that some
parameters, not connected to the Green's function under analysis,
appear in the expressions: $\chi _c^\alpha = \langle n(i)n^\alpha
(i) \rangle$ and  $p = {1 \over 4} \langle n_\mu ^\alpha (i)n_\mu
(i)\rangle - \langle [c_\uparrow (i)c_\downarrow (i)]^\alpha
c_\downarrow ^\dagger (i)c_\uparrow ^\dagger (i) \rangle$. The
first one, $\chi _c^\alpha$, will be computed by calculating the
density-density correlation function $\langle n(i) n(j) \rangle$
within the one-loop approximation \cite{Mancini:95b} (see second
equation in Eqs.~(\ref{Pauli})). The second will be fixed by the
local algebra constraint \cite{Mancini:00} $\langle \xi(i)
\eta^\dagger(i) \rangle=0$. Such a procedure is very peculiar to
the COM treatment \cite{Mancini:00} as discussed in the
introduction. By solving the set of coupled self-consistent
equations (\ref{Eq.5.1.11}-\ref{Pauli}), we can calculate the
various correlation functions and the physical properties of the
system. Results will be presented in the following sections for
the one-dimensional (1D) and the two-dimensional (2D) systems.

Before moving to the results, it is worth noting that the set of
self-consistent equations (\ref{Eq.5.1.11}-\ref{Pauli}) is highly
non-linear. According to this, it is natural to expect a certain
number of coexisting solutions with quite different features.
Actually, the set admits only two distinct solutions that,
hereafter, we will call COM1 and COM2, according to the main sign
of parameter $p$. In particular, as a function of the filling, we
have a $p$ positive and of the order of the filling in COM1 and a
$p$ negative or very small and positive in COM2. As the Composite
Operator Method tries to give answers in the whole space of model
and physical parameters and as the Hubbard model response is
profoundly different according to the region of the latter space,
the presence of two solutions, so different in their features,
should be seen as a richness of the method. Due to the
difficulties inherent to the task of studying the whole phase
diagram, many other approximations focus just on one region and
usually give wrong results in the rest of the parameter space.
COM, also thanks to the opportunity of choosing between the two
solutions according to the features one expects to describe, has
proved to be capable to explore the whole phase diagram, to get
significatively good results for all dimensions and values of the
tuning parameters, to give relevant interpretations of the
experimental facts and to constitute a solid basis for further
investigations. It is also worth remembering that, at the end of
the day, our analysis has, as main goal, the description of
features, more or less anomalous, of real physical systems by
means of Hamiltonian models. We can never be sure that our
Hamiltonian contains all, and only, relevant ingredients.
Different experimental situations can be described by different
analytical (and/or numerical) solutions of the same Hamiltonian.
For a fixed set of parameters (filling, temperature, pressure,
...), the different solutions will have different free energies
and one can simply think to choose one or the other according to
this. This is the correct procedure if we wish to answer the
question: "Which is the phase effectively described by this
Hamiltonian under such external conditions?". On the other hand, a
solution different from the one with lower free energy can better
describe the real experimental situation. This latter, although
mainly determined by the ingredients already present in the
Hamiltonian (the possibility to find a solution with similar
characteristics assures this), is that effectively realized in
nature owing to some marginal interactions not present in the
Hamiltonian chosen. If we would include such interactions in the
Hamiltonian and compute new solutions, we will find that the lower
free energy one will be the one capable to describe the actual
physical situation. Now, the most important consideration to be
done is that this latter solution will be essential identical to
that obtained with the previous Hamiltonian and that was already
capable to describe the actual physical situation although it has
not the lower free energy. According to this, in the comparison
with the experiments no solution can be discarded a priori
(neither on the basis of the free energy determination) and, with
this idea in mind, we have here, and in many other works,
presented the results for the two solutions we have obtained.

For the one-dimensional system we will consider only COM2 solution
as our past experience suggests that this is the one best suited
to describe the physics of the simple one-dimensional Hubbard
model for which we got excellent agreements with the Bethe Ansatz
exact solution \cite{Avella:98b,Avella:98e,Sanchez:99,Avella:00}.
For the two-dimensional case, we will study both solutions as they
will permit us to analyze, in particular in proximity of half
filling and as regards the metal-insulator transition and its
relation to the charge ordering transition, two different
behaviors that could be both observed experimentally. Also in this
case the many positive comparisons with numerical results, that we
have obtained in the previous studies on the simple
two-dimensional Hubbard model
\cite{Mancini:95,Mancini:95a,Mancini:95b,Avella:98,Avella:98f,Mancini:99a},
will be used as a guide throughout all the analysis.

\section{The Chemical Potential}

The chemical potential can be determined as a function of the
parameters $n$, $T$, $U$, $V$ by solving the system of
self-consistent equations (\ref{Eq.5.1.11}-\ref{Pauli}). It is
worth noticing that our results show that the relation
\begin{equation}\label{Eq.5.2.1}
\mu (2-n)=U+4dV-\mu (n)
\end{equation}
required by the particle-hole symmetry, is exactly satisfied. In
particular at half filling we have $\mu (1)={1 \over 2}U+2dV$.
This is due to the fact that among the possible representations
the Algebra Constraint coming from the Pauli principle [see first
equation in Eqs.~(\ref{Pauli})] selects the one which preserves
the particle-hole symmetry of the model \cite{Mancini:00}. Any
other choice for the equation fixing the parameter $p$ leads to a
violation of the symmetry.

\subsection{One-dimensional system}

\begin{figure}[tbp]
\begin{center}
\includegraphics[width=7.5cm,clip=]{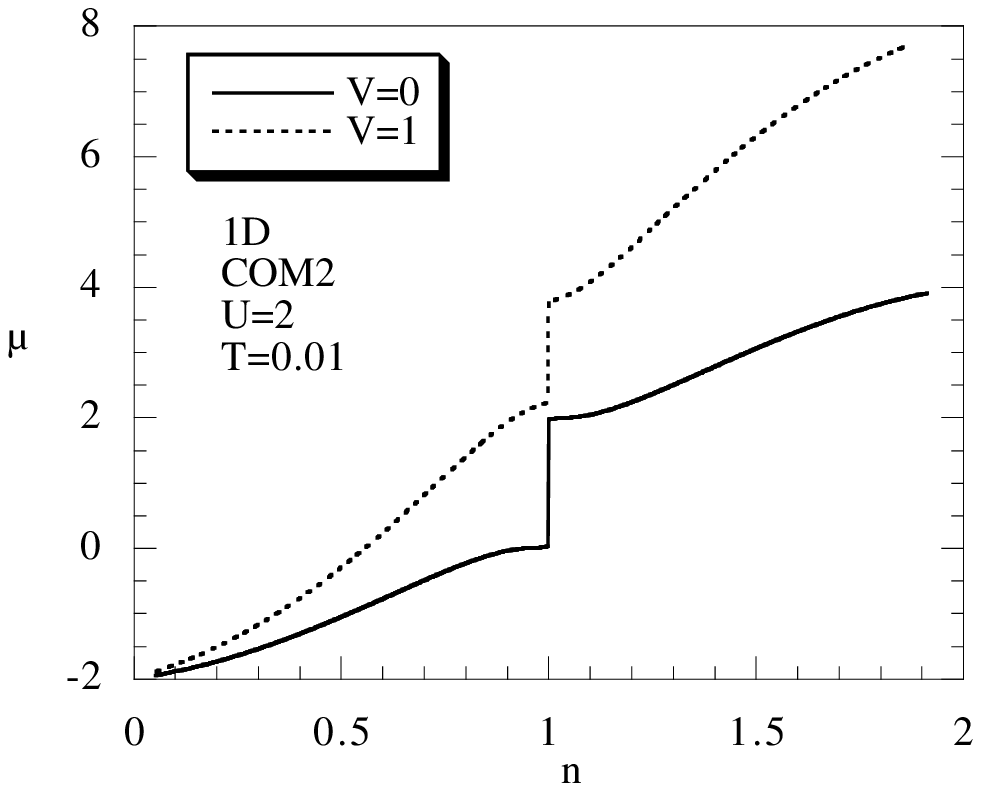}
\hfill
\includegraphics[width=7.5cm,clip=]{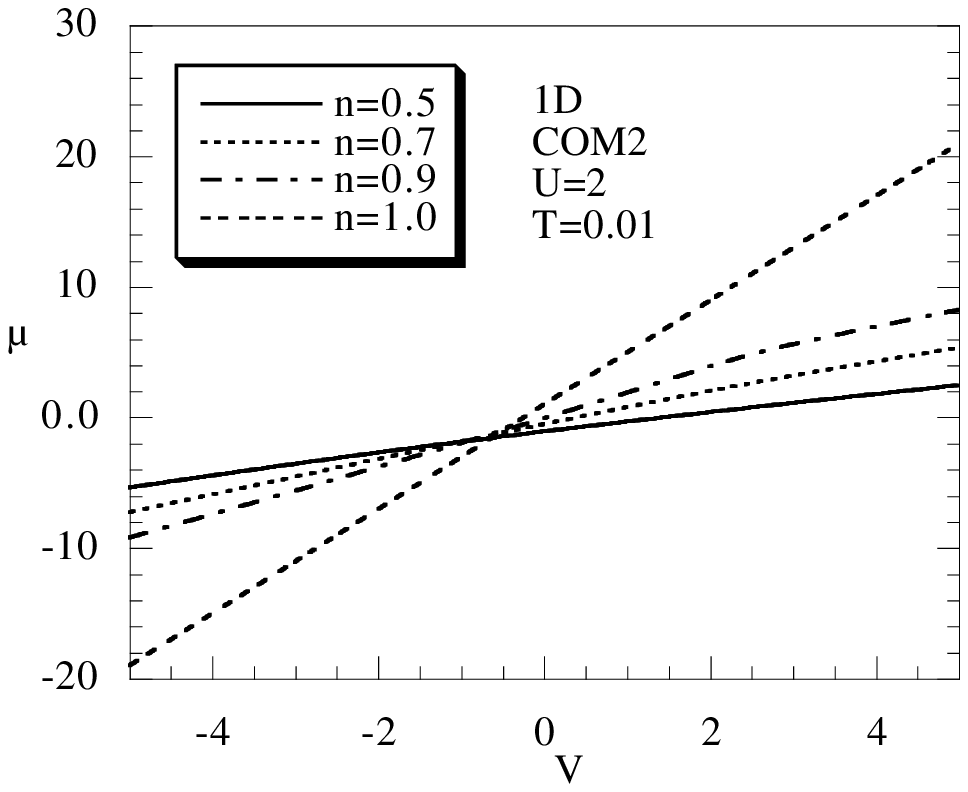}
\end{center}
\caption{The chemical potential at $U=2$ as a function of: (top)
the particle density for $V=0$, $1$; (bottom) the intersite
Coulomb interaction for several values of $n$.} \label{Fig.5.2.1}
\end{figure}

The chemical potential as a function of the particle density is
reported in Fig.~\ref{Fig.5.2.1} (top) for $U=2$ and $V=0$ and
$1$. For any value of the filling the chemical potential increases
by increasing the intersite Coulomb interaction and the increment
is an increasing function of the filling: within the paramagnetic
phase the average probability for two particles to be nearest
neighbors increases with the filling and accordingly increases the
free energy in the presence of a repulsive intersite Coulomb
interaction, then, the behavior of the chemical potential follows.

For negative (i.e., attractive) values of the intersite Coulomb
potential, we can see that the paramagnetic solution is unstable
towards a phase with charge separation (i.e., with charge
ordering). As a matter of fact, an attractive potential between
charges at nearest neighbor sites favors a rearrangement of the
particles in a one-particle-per-site scheme in order to maximize
the gain in energy. At fillings less than one this scheme can
lower the energy more and more by accepting more particle in the
system and increasing the number of occupied couples of nearest
neighbor sites. According to this, the chemical potential shows a
negative slope which, on increasing the value of the intersite
potential, manifests at lower and lower values of the filling. The
critical value of the intersite Coulomb potential which controls
the transition to the charge ordered state depends on the
intensity of the local Coulomb interaction. In order to illustrate
the case, in Fig.~\ref{Fig.5.2.1} (bottom), the chemical potential
is given versus intersite Coulomb potential for $U=2$ and various
values of the particle density: all curves cross at a certain
value $V \approx -0.5$, below which the system exhibits a negative
compressibility.

\begin{figure}[tbp]
\begin{center}
\includegraphics[width=7.5cm,clip=]{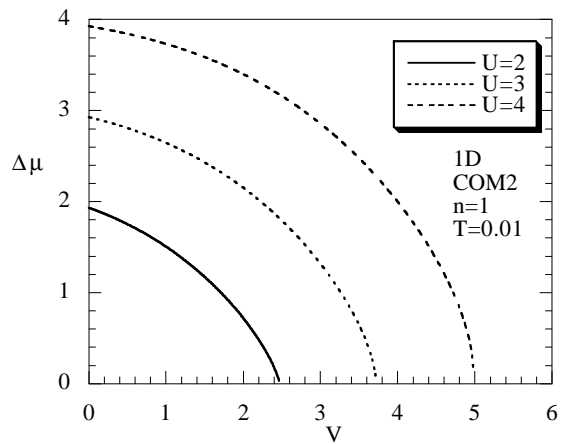}
\end{center}
\caption{The discontinuity of the chemical potential at half
filling and $T=0.01$ is plotted versus the intersite Coulomb
interaction for $U=2$, $3$ and $4$.} \label{Fig.5.2.3}
\end{figure}

As expected from the exact Bethe Ansatz solution of the
one-dimensional simple Hubbard model, for any $U>0$ there is a
discontinuity at half filling, signalling a gap in the density of
states and the occurrence of an insulating phase. As it can be
seen from Fig.~\ref{Fig.5.2.1} (top), the effect of the intersite
term $V$ is to reduce the size of the gap. This is studied in
Fig.~\ref{Fig.5.2.3} where the discontinuity of the chemical
potential at half filling
\begin{equation}\label{Eq.5.2.3}
\Delta \mu =\mu _+(1)-\mu _-(1)
\end{equation}
is plotted versus the intersite Coulomb potential for various
values of the onsite Coulomb potential at half filling and
$T=0.01$. We see that for a given value of the onsite Coulomb
potential, there is a critical value $V_c$. For values of
intersite Coulomb potential greater than $V_c$, the paramagnetic
insulating phase becomes unstable (the chemical potential gets a
negative slope) and there is a phase transition to a charge
ordered insulating state (CO) \cite{Hirsch:84a}. This kind of
ordering is much different than that discussed previously (i.e.,
the one-particle-per-site type); in this case the repulsion among
particles favors a checkerboard pattern with half sites double
occupied (see in the next sections the discussion about the double
occupancy) and half empty, these latter ones being the nearest
neighbors of the former ones. By analyzing the derivative
$[\partial (\Delta \mu )/\partial V]_{V=V_c}$, we can show that at
zero temperature the transition is second order for $U \le 2$ and
first order for higher values of the onsite Coulomb potential.


\subsection{Two-dimensional system}

\begin{figure}[tbp]
\begin{center}
\includegraphics[width=7.5cm,clip=]{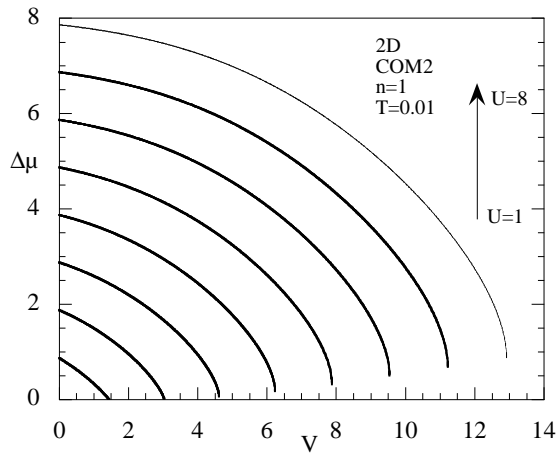}
\end{center}
\caption{The discontinuity of the chemical potential at half
filling is plotted versus the intersite Coulomb interaction for
various values of $U$ and $T=0.01$.} \label{Fig.5.2.4}
\end{figure}

\begin{figure}[tbp]
\begin{center}
\includegraphics[width=7.5cm,clip=]{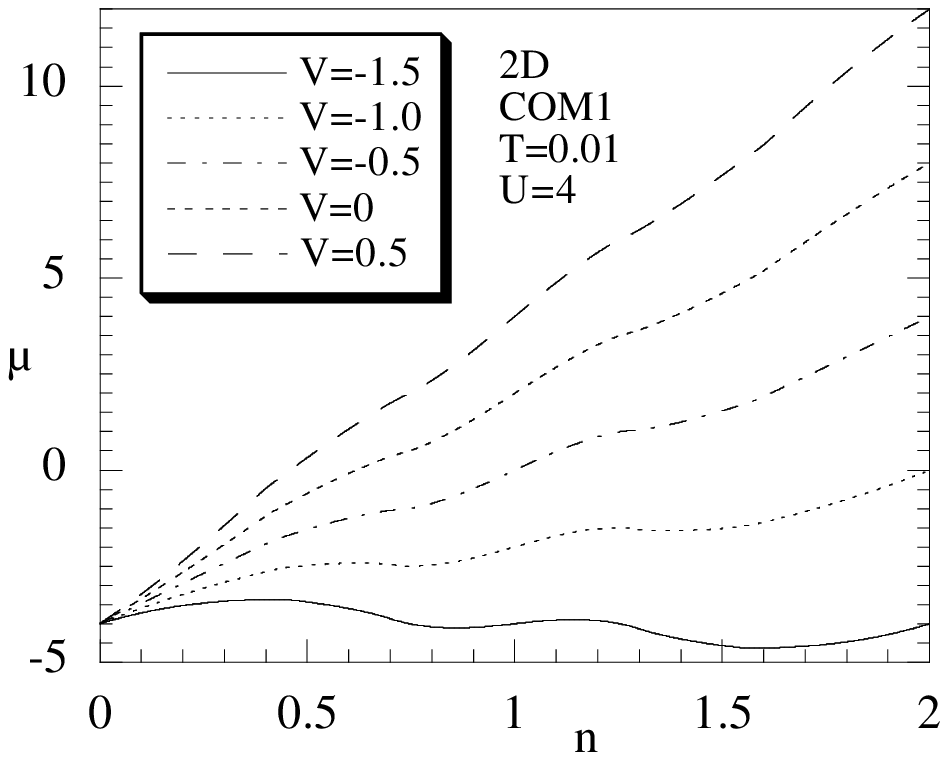}
\hfill
\includegraphics[width=7.5cm,clip=]{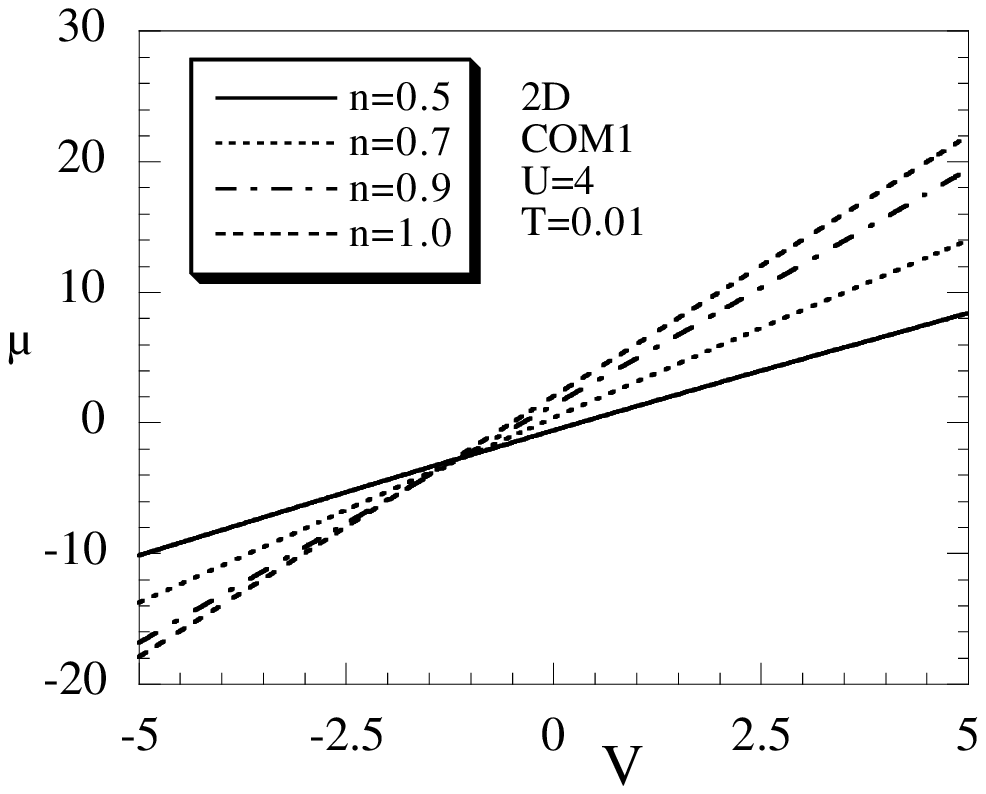}
\end{center}
\caption{The chemical potential at $T=0.01$ and $U=4$ versus:
(top) the particle density for several values of $V$; (bottom) the
intersite interaction for several values of $n$.}
\label{Fig.5.2.5}
\end{figure}

\begin{figure}[tbp]
\begin{center}
\includegraphics[width=7.5cm,clip=]{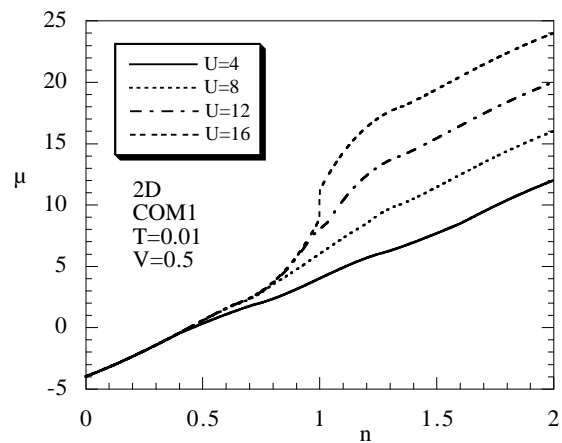}
\end{center}
\caption{The chemical potential versus the particle density at
$T=0.01$, $V=0.5$ and several values of $U$.} \label{Fig.5.2.6}
\end{figure}

As already noted for the 1D system, also in 2D within COM2
solution the discontinuity $\Delta \mu $ of the chemical potential
at half filling decreases by increasing the value of the intersite
Coulomb potential and the system exhibits a phase transition to a
charge ordered phase at some critical value $V_c$. The values of
$V_c$, as a function of the onsite Coulomb potential, are shown in
Fig.~\ref{Fig.5.2.4} where $\Delta \mu $ is reported versus the
intersite Coulomb potential for various values of the onsite
interaction $U$.

Within COM1 solution, the behavior of the chemical potential as a
function of the particle density is shown in Fig.~\ref{Fig.5.2.5}
(top) for $U=4$ and various values of the intersite Coulomb
potential. We see that for attractive values of the intersite
Coulomb interaction the chemical potential decreases by increasing
the filling, showing an instability of the paramagnetic case
towards phase separation. In particular, exactly at quarter
filling (n=0.5), we can have a charge-ordered state of
one-particle-per-site type. Away from quarter filling, the phase
separation is between two phases with different particle densities
(these latter should be determined through a Maxwell
construction), whose nature could be investigated (they can be
both charge-disordered or one ordered and another disordered)
within a treatment where translational invariance should be
relaxed. The instability of the paramagnetic phase can be here
studied by plotting the chemical potential as a function of the
intersite Coulomb potential. This is shown in Fig.~\ref{Fig.5.2.5}
(bottom) where we see that the curves for different values of
filling all cross at some critical value $V^*$, depending on the
onsite Coulomb potential [$V^*\approx -1.1$ for $U=4$]. For
$\left| V \right|<\left| {V^*} \right|$ the chemical potential is
a decreasing function of the filling.

As in the simple Hubbard model
\cite{Mancini:95,Mancini:95a,Mancini:95b,Avella:98,Avella:98f,Mancini:99a},
COM1 solution describes a metallic phase in the low regime of
on-site Coulomb repulsion and a transition to the insulating state
when the potential reaches some critical value. This feature
survives also when the intersite Coulomb interaction is taken into
account. For a fixed value of this latter, there is a critical
value of the on-site repulsion such that the chemical potential
exhibits a discontinuity at half filling. This latter signals the
opening of a gap in the density of states and therefore a Mott
transition. This feature is illustrated in Fig.~\ref{Fig.5.2.6}.
Now, the current study shows that the critical value $U_c$
decreases by increasing the intersite potential $V$. By further
increasing the intersite potential, the system undergoes a second
transition to a charge ordered state of checkerboard type. This
transition is characterized by a discontinuity in the double
occupancy as commented above and shown in the next sections.

\section{Phase diagrams}

\subsection{One-dimensional system}

\begin{figure}[tbp]
\begin{center}
\includegraphics[width=7.5cm,clip=]{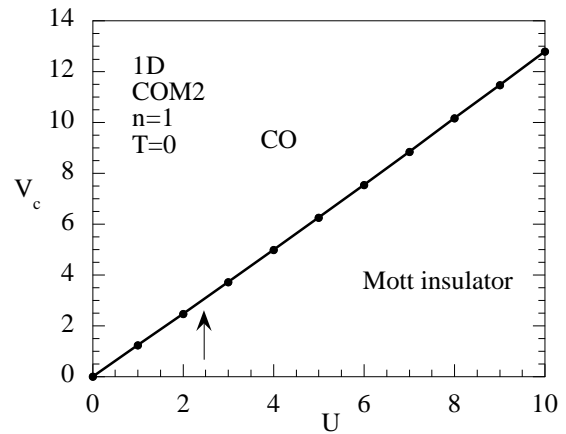}
\end{center}
\caption{The critical value $V_c$ where there is a phase
transition from the Mott insulator to charge ordered state is
plotted as a function of the onsite Coulomb potential at half
filling and zero temperature.} \label{Fig.5.3.1}
\end{figure}

As reported in the previous section, COM2 solution for the one
dimensional system shows that there is phase transition from the
Mott insulating phase to a inhomogeneous charge ordered state of
checkerboard type for positive values of the intersite Coulomb
potential greater than some $V_c$. This result is consistent with
many other studies
\cite{Hirsch:84a,Cannon:91,vanDongen:94a,Japaridze:99,Nakamura:00,Tsuchiizu:02,Sengupta:02,Jeckelmann:02}.
The nature of this phase transition is not well understood yet and
currently under intense
investigation\cite{Nakamura:99,Nakamura:00,Tsuchiizu:02,Sengupta:02,Jeckelmann:02}.
The phase diagram in the plane $V$-$U$ is shown in
Fig.~\ref{Fig.5.3.1}, where the critical value $V_c$ of the
intersite Coulomb potential is plotted as a function of the onsite
potential $U$. The arrow indicates the point where the phase
transition from second order changes to first order.

It is necessary noticing that this transition is quite different
from that usually observed in proximity of the $U=2V$ line between
an inhomogeneous charge ordered phase and an homogenous spin
ordered phase. In this manuscript, we decided to focus on the
homogenous paramagnetic phase, its instabilities towards charge
ordered inhomogeneous phases (paying attention to the rank of the
transition) and the relation between these latter and the
metal-insulator transition. These kinds of transitions are the
only ones relevant at room temperatures and/or in presence of
frustration as the antiferromagnetic phase is quite depressed in
such cases.

\begin{figure}[tbp]
\begin{center}
\includegraphics[width=7.5cm,clip=]{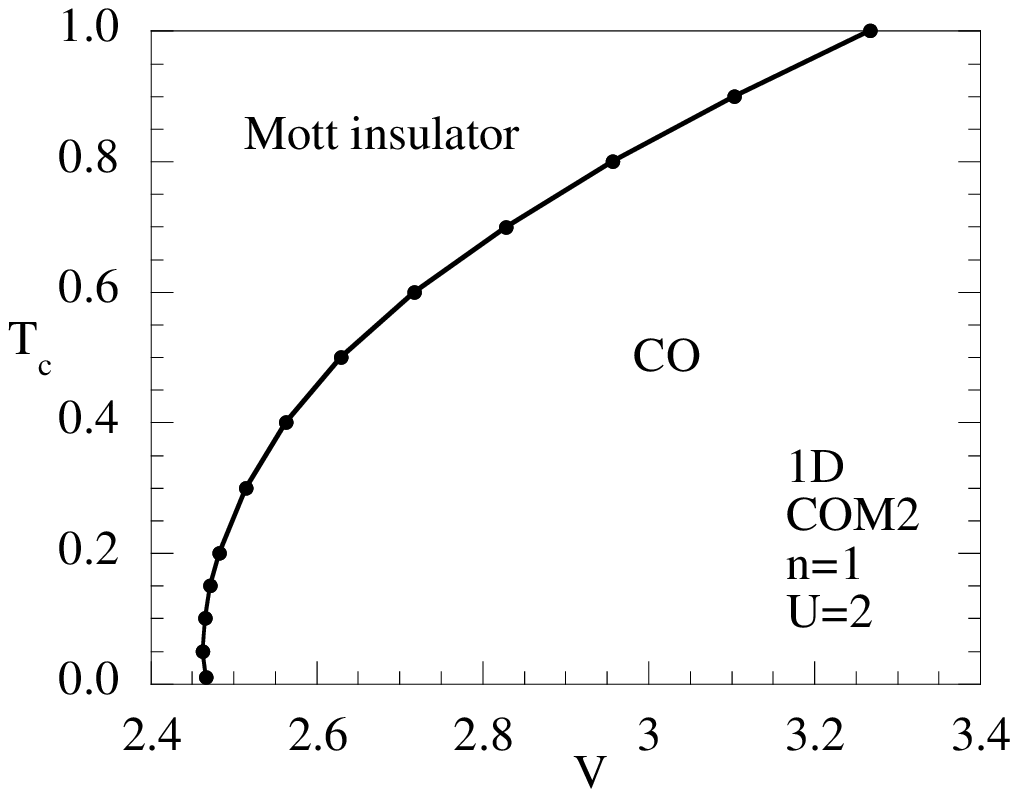}
\hfill
\includegraphics[width=7.5cm,clip=]{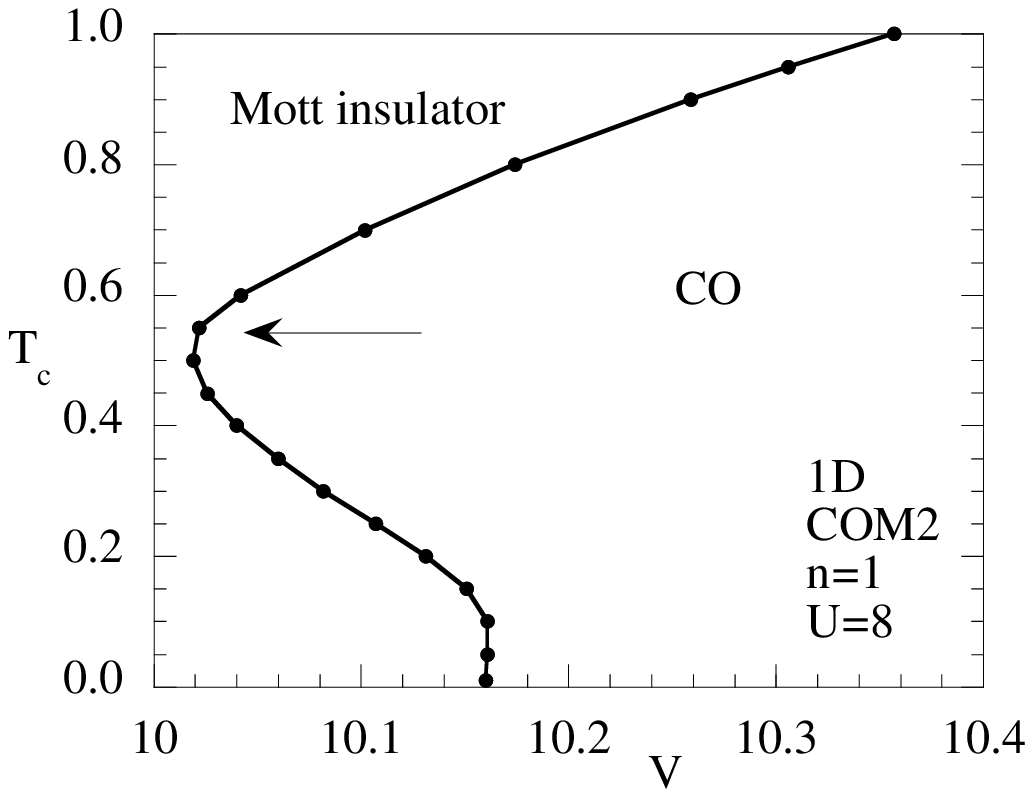}
\end{center}
\caption{The critical temperature $T_c$ for the Mott insulator to
charge ordered state phase transition is plotted as a function of
the intersite Coulomb potential at half filling and $U=2$ (top)
and $U=8$ (bottom).} \label{Fig.5.3.2}
\end{figure}

\begin{figure}[tbp]
\begin{center}
\includegraphics[width=7.5cm,clip=]{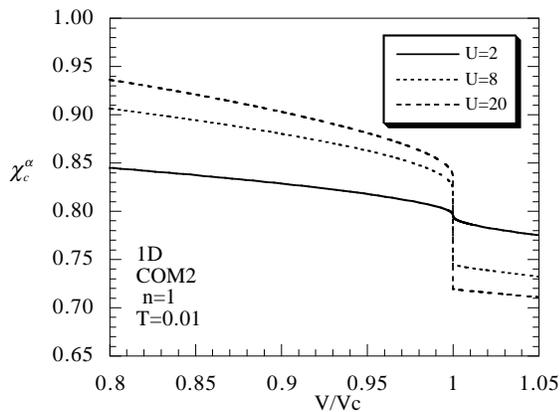}
\end{center}
\caption{The nearest-neighbor density-density correlation function
$\chi^\alpha_c$ is plotted as a function of $V/V_c$ at half
filling, $T=0.01$ and $U=2$, $8$ and $20$.} \label{Fig.5.3.3}
\end{figure}

In Fig.~\ref{Fig.5.3.2} we give the phase diagram in the plane
$T$-$V$ for $U=2$ and $U=8$. For $U=2$ the transition is second
order for all values of $T$. For $U=8$ the transition is first
order for $T\le 0.55$ and second order for $T\ge 0.6$. The arrow
indicates the temperature where, by increasing $T$, the transition
becomes continuous. It is interesting to observe that the
transition is continuous if there is no reentrant temperature
behavior. By this latter we mean a situation in which by
increasing temperature we can first enter and then exit a phase
when within another (e.g., see Fig.~\ref{Fig.5.3.2} (bottom) at
$V=10.1$ for increasing temperature from zero). When there is a
reentrant temperature behavior, the transition is discontinuous up
to the turning point, then becomes continuous. A reentrant
temperature behavior has been experimentally observed
\cite{Tomioka:97,Chatterji:00,Dho:01} and will be further
discussed for the 2D system. It is worth noting that the
transition is also marked by a discontinuity in the
nearest-neighbor density-density correlation function $\chi
_c^\alpha = \langle n^\alpha (i)n(i) \rangle$. This quantity is
shown in Fig.~\ref{Fig.5.3.3} as a function of $V/V_c$ at half
filling, $T=0.01$ and $U=2$, $8$ and $20$. At the transition $\chi
_c^\alpha $ is continuous for $U=2$ and discontinuous for $U>2$.
The nearest-neighbor density-density correlation function is a
decreasing function of the intersite potential as the repulsion
diminishes the probability of finding neighboring sites occupied.
In particular, at the transition this probability reduces with a
change of concavity (second order transition) or with a
discontinuity (first order transition).

\begin{figure}[tbp]
\begin{center}
\includegraphics[width=7.5cm,clip=]{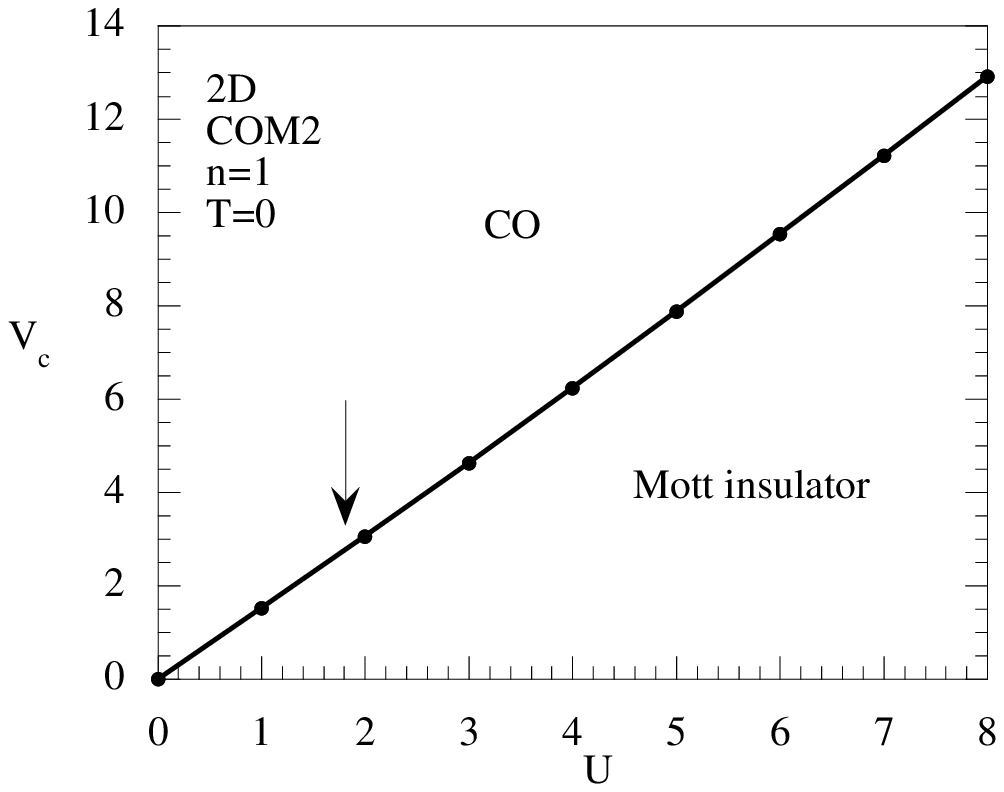}
\hfill
\includegraphics[width=7.5cm,clip=]{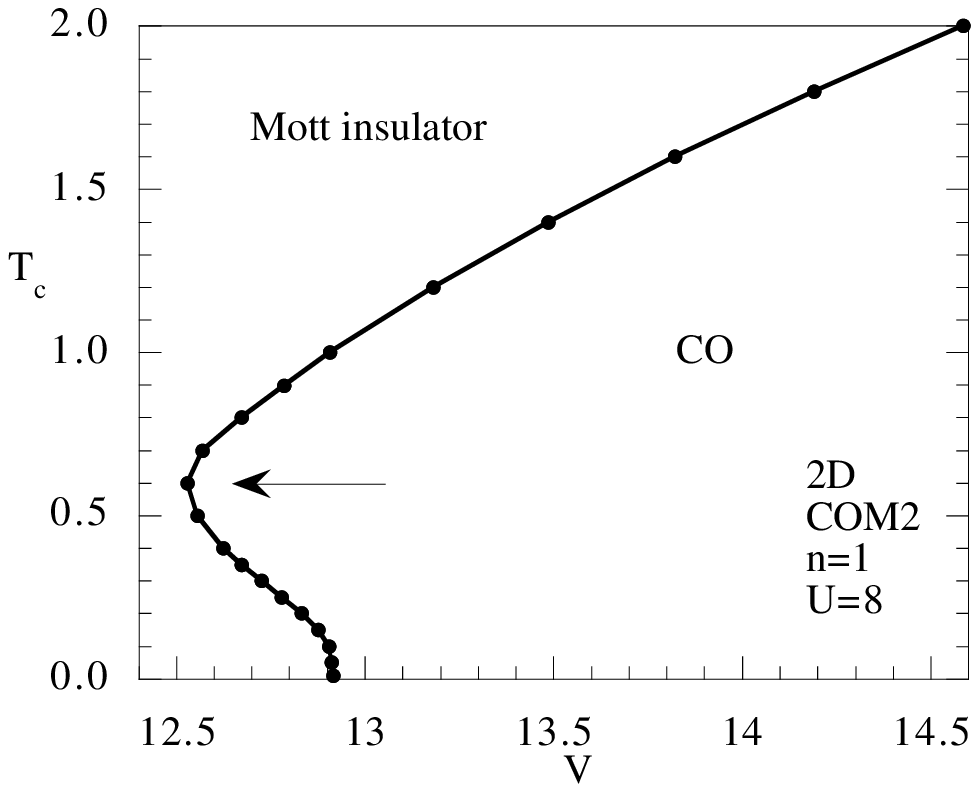}
\end{center}
\caption{(top) The critical value $V_c$ for the Mott insulator to
charge ordered state phase transition is plotted for the
two-dimensional case (COM2 solution) as a function of: (top) the
onsite Coulomb potential at half filling and zero temperature
(COM2 solution); (bottom) $V$ at half filling and $U=8$.}
\label{Fig.5.3.4}
\end{figure}

\begin{figure}[tbp]
\begin{center}
\includegraphics[width=7.5cm,clip=]{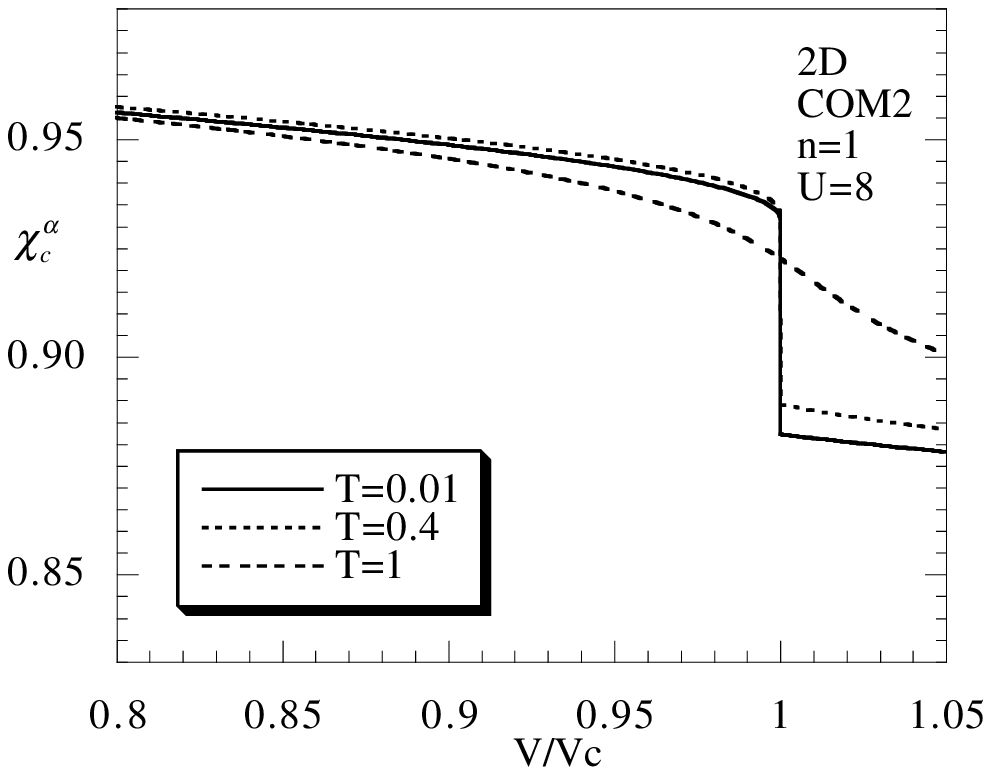}
\hfill
\includegraphics[width=7.5cm,clip=]{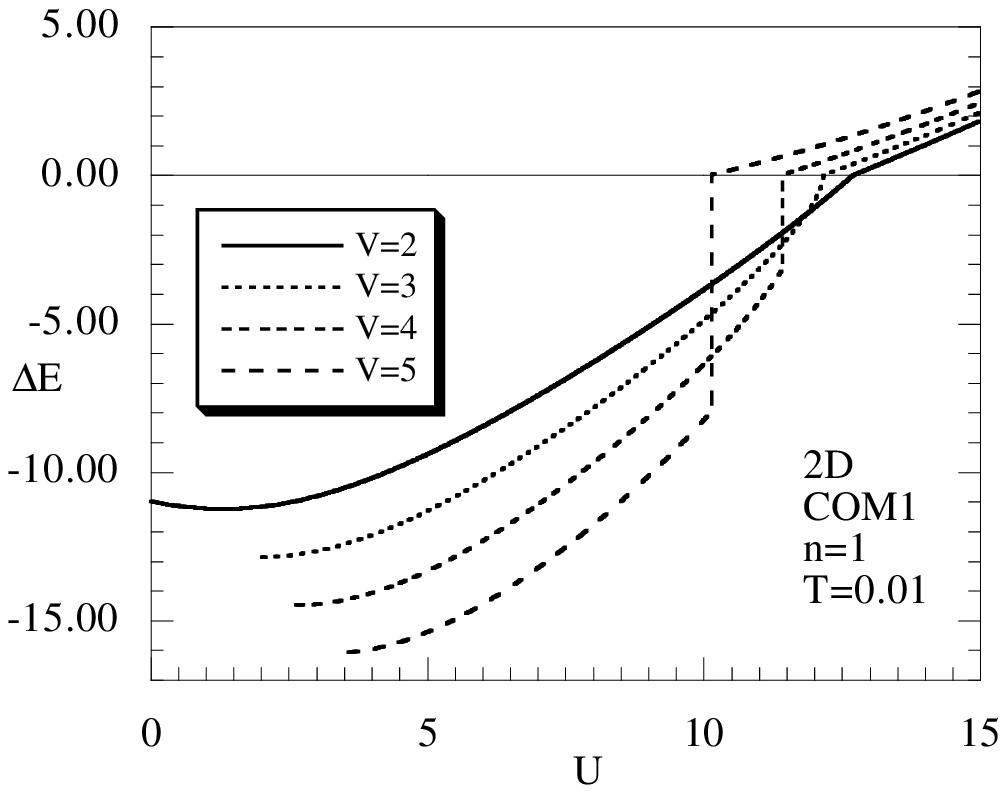}
\end{center}
\caption{(top) The nearest-neighbor density-density correlation
function $\chi^\alpha_c$ is plotted as a function of $V/V_c$ at
half filling, $U=8$ and $T=0.01$, $0.4$ and $T=1$, for the
two-dimensional case (COM2 solution). (bottom) The energy gap
$\Delta E$ at half filling is reported as a function of U for
several values of U for T=0, for the two-dimensional case (COM1
solution).} \label{Fig.5.3.5}
\end{figure}

\subsection{Two-dimensional system: COM2}

The COM2 solution for the two-dimensional case has similar
characteristic to the 1D case. The phase diagrams in the planes
$V$-$U$ and $T$-$V$ are shown in Figs.~\ref{Fig.5.3.4} (top) and
~\ref{Fig.5.3.4} (bottom), respectively.

At zero temperature the transition is continuous for $U\le 1.8$
and first order for $U>1.9$. For finite temperature and $U=8$ a
reentrant behavior as function of temperature is observed. The
transition is first order up to the turning point $T=0.6$, then
becomes continuous. For $U\le 1.8$ no re-entrant behavior is
observed. The fact that charge ordering may disappear by
decreasing temperature has been experimentally observed in $Pr
_{0.65}(Ca_{0.7}Sr_{0.3})_{0.35}MnO_3$ \cite{Tomioka:97} and
$La_{2-2x}Sr_{1+2x}Mn_2O_7$ $(0.47\le x\le 0.62)$
\cite{Chatterji:00,Dho:01}. On the theoretical side a re-entrant
temperature has been obtained in the context of the extended
Hubbard model at quarter filling
\cite{Pietig:99,Hellberg:01,Hoang:02}.

In Fig.~\ref{Fig.5.3.5} (top) the nearest-neighbor density-density
correlation function $\chi _c^\alpha = \langle n^\alpha (i)n(i)
\rangle$ is plotted as a function of $V/V_c$ at half filling,
$U=8$ and various temperatures. At the transition $\chi _c^\alpha
$ is discontinuous for $T=0.01$ and $T=0.4$, and continuous for
$T=1$, in agreement with the phase diagram shown in
Fig.~\ref{Fig.5.3.4} (bottom).

\subsection{Two-dimensional system: COM1}

The results given for the chemical potential show that for low
values of the on-site Coulomb interaction the system is in a
metallic state and undergoes a metal-insulator transition for a
critical value $U_c$, which depends on the intensity of the
intersite potential. In order to study this metal-insulator
transition (MIT) we consider the quantity
\begin{equation}\label{Eq.5.3.1}
\Delta E=E_1(0,0)-E_2(\pi ,\pi )
\end{equation}
which measures the distance between the bottom of the upper
subbands and the top of the lower one. After Eq.~(\ref{Eq.5.1.7}),
at half filling we have
\begin{alignat}{2}\label{Eq.5.3.2}
& R({\bf k})=R_1\alpha ({\bf k}) & \quad &
R_1=-8tp+4V(C_{22}^\alpha +C_{11}^\alpha ) \nonumber \\
& g({\bf k})=g_0 & & g_0=-U+8V(1-\chi _c^\alpha ) \\
& m_{12}({\bf k})=m_1\alpha ({\bf k}) & & m_1=-4 \left[ t({1 \over
2}-p)-V C_{12}^\alpha \right] \nonumber
\end{alignat}
Therefore, we simply have
\begin{equation}\label{Eq.5.3.3}
\Delta E=2R_1+\sqrt {g_0^2+16m_1^2}
\end{equation}

For a fixed value of the onsite Coulomb repulsion, the equation
$\Delta E=0$ will determine the critical value $V_c$ where a
metal-insulator transition occurs. The results show that the
metallic region is compressed by the presence of the intersite
interaction and disappears for $V>5.7$. This is shown in
Fig.~\ref{Fig.5.3.5} (bottom), where $\Delta E$ is reported versus
the onsite Coulomb potential for various values of the intersite
Coulomb potential at zero temperature.

\begin{figure}[tbp]
\begin{center}
\includegraphics[width=7.5cm,clip=]{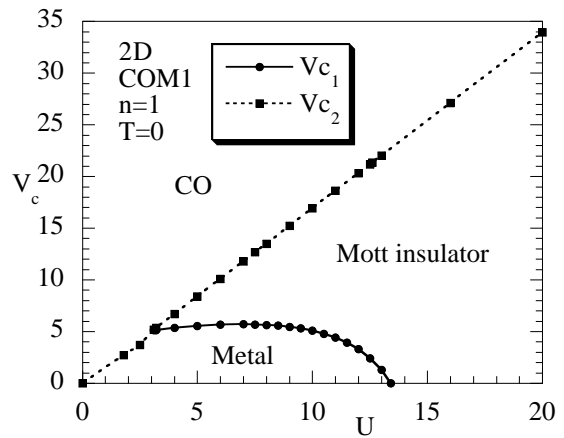}
\end{center}
\caption{The complete phase diagram in the plane $U$-$V$ is shown
at half filling for zero temperature, for the two-dimensional case
(COM1 solution).} \label{Fig.5.3.7}
\end{figure}

\begin{figure}[tbp]
\begin{center}
\includegraphics[width=7.5cm,clip=]{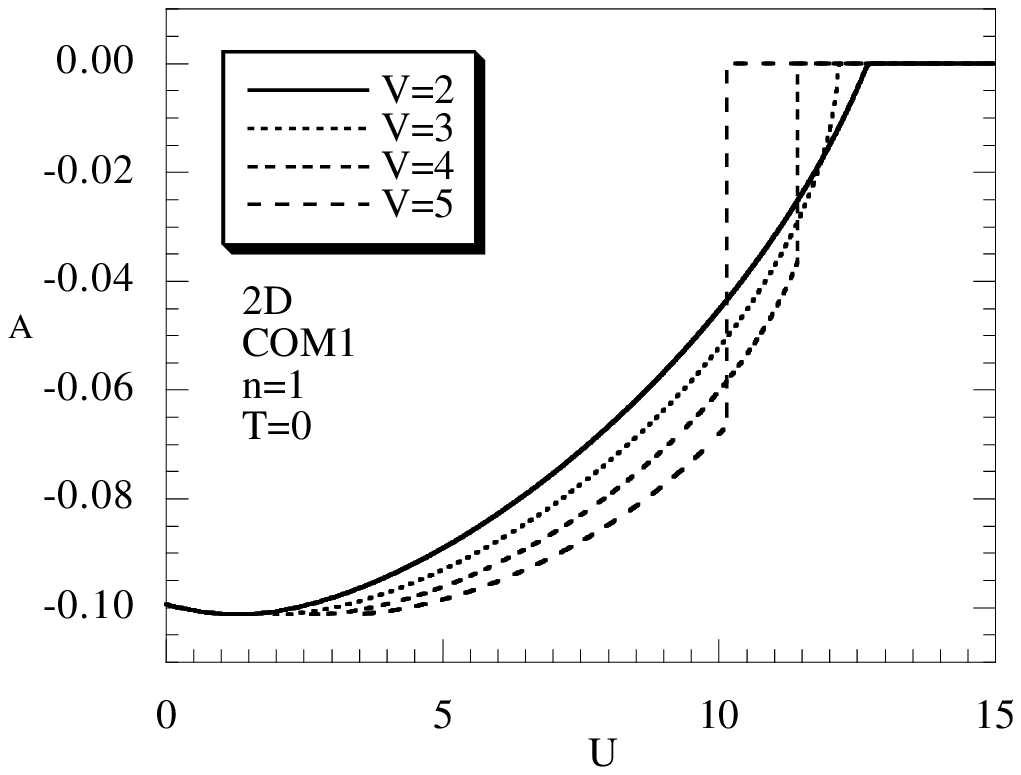}
\hfill
\includegraphics[width=7.5cm,clip=]{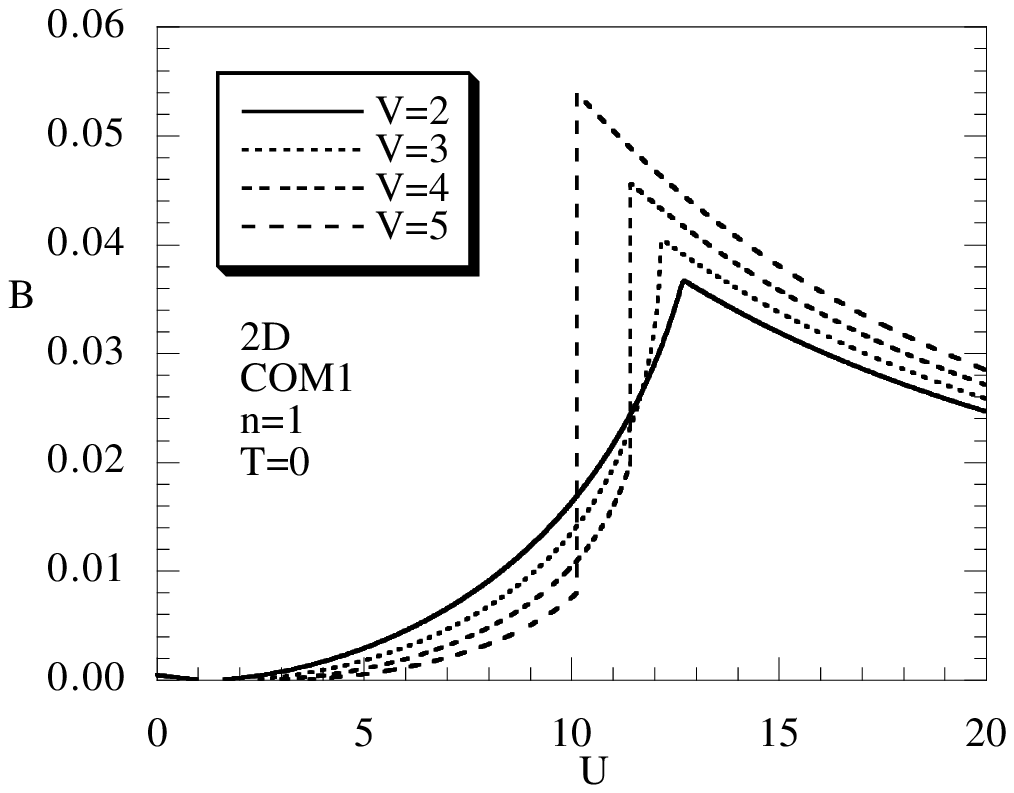}
\end{center}
\caption{The nearest neighbor hopping amplitudes $A$ and $B$ are
reported as functions of the onsite Coulomb potential for various
values of $V$, at half filling and $T=0$.} \label{Fig.5.3.8}
\end{figure}

\begin{figure}[tbp]
\begin{center}
\includegraphics[width=7.5cm,clip=]{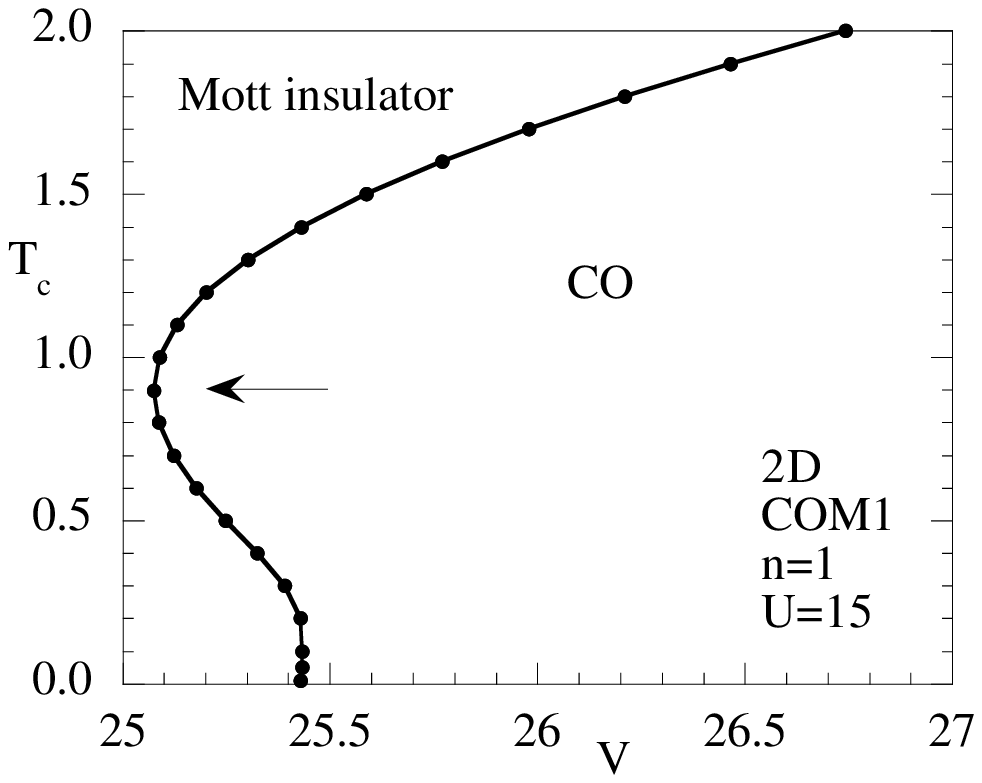}
\hfill
\includegraphics[width=7.5cm,clip=]{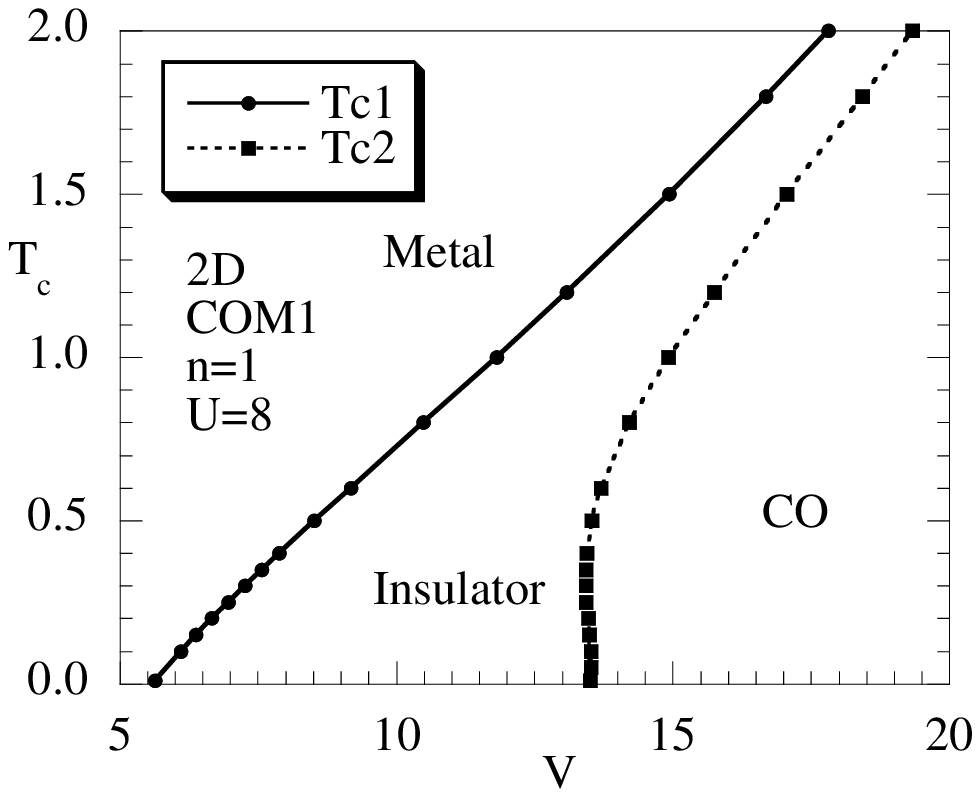}
\end{center}
\caption{The phase diagram in the plane (V-T) is given for $U=15$
(top) and $U=8$ (bottom).} \label{Fig.5.3.9}
\end{figure}

For $V>5.7$ the equation $\Delta E=0$ does not have a solution and
the metallic phase disappears. By further increasing the intersite
Coulomb potential, there is an instability in the self-consistent
equations (\ref{Eq.5.1.11}-\ref{Pauli}), the double occupancy
exhibits a discontinuity leading the system to a charge ordered
state. The complete phase diagram in the plane $V$-$U$ is shown in
Fig.~\ref{Fig.5.3.7}. The diagram is characterized by two critical
curves, $V_{c1}$ and $V_{c2}$, which separate the different
phases. $V_{c1}$ controls the MIT and $V_{c2}$ controls the
transition to a charge ordered state. At $V_{c1}$ the transition
is first order for $U \le 12$ and second order for $U\ge 12.2$; at
$V_{c2}$ we have first order for $U\ge 1.9$ and second order for
$U\le 1.8$. As proposed in Ref.~\onlinecite{Mancini:00b}, the
insulating state can be characterized by the order parameters
$\langle \xi_\sigma (j)\xi_\sigma^\dagger (j_{odd})\rangle$ and
$\langle \eta_\sigma (j)\eta_\sigma^\dagger (j_{odd}) \rangle$
which vanish at the MIT ($j_{odd}$ is any site reachable in an odd
number of hops from site $j$). The nearest neighbor hopping
amplitudes $A=\langle \xi ^\alpha (i)\xi^\dagger (i)\rangle
=\langle \eta ^\alpha (i)\eta ^\dagger (i)\rangle$ and
$B=\langle\eta ^\alpha (i)\xi^\dagger (i)\rangle = \langle \xi
^\alpha (i)\eta ^\dagger (i)\rangle$ are reported in
Figs.~\ref{Fig.5.3.8} as functions of the onsite Coulomb potential
for various values of the intersite Coulomb potential. We see that
for a given value of the intersite Coulomb potential, the
probability amplitude $A$ suddenly vanishes at some critical value
of $U_c$ and remains zero for all values of $U>U_c$. The
probability amplitude $B$ does not vanish above $U_c$. Owing to
this contribution, we have that for $U>U_c$ the hopping of
electrons from site $j$ to a nearest neighbor is not forbidden,
although restricted by the fact that $A=0$. This is consistent
with the fact that the double occupancy remains finite for $U>U_c$
and vanishes only in the limit $U\to \infty$ \cite{Mancini:00b}.

The phase diagram in the plane $T$-$V$ is shown in
Figs.~\ref{Fig.5.3.9} (top) and \ref{Fig.5.3.9} (bottom). For
$U=15$ (see Fig.~\ref{Fig.5.3.9} (top)) there is no metallic phase
and we have a critical temperature $T_c(V)$ where a transition
from insulating to charge ordered state is observed. The
transition is first order up to $T_c=0.95$, then becomes
continuous. A re-entrant temperature is observed with the same
characteristics previously discussed. For $U=8$ (see
Fig.~\ref{Fig.5.3.9} (bottom)) we have two critical temperatures,
$T_{c1}$ and $T_{c2}$, which characterize the MIT transition and
the insulator-charge order transition, respectively. Also in this
case a re-entrant temperature is observed in the latter
transition.

\section{Double Occupancy, Kinetic and Internal Energies}

In contrast to the local interaction $U$, a positive intersite
interaction $V$ favors the double occupancy $D=\frac1N \sum_{\bf
i} \langle n_\uparrow({\bf i}) n_\downarrow({\bf i}) \rangle =
\frac n2- C_{22}$. This happens through the mechanism which favors
the formation of the checkerboard type charge ordered state as
discussed above. The cost in energy $U$ of having a double
occupied site is partly balanced by the cost in energy $V$ related
to the presence of two particles on nearest-neighbor sites. On the
contrary, an attractive intersite Coulomb interaction $V$ leads to
a decrement of the double occupancy as the one-particle-per-site
type of charge ordering favors the presence of single occupied
sites with respect to double occupied ones.

\subsection{One-dimensional system}

\begin{figure}[tbp]
\begin{center}
\includegraphics[width=7.5cm,clip=]{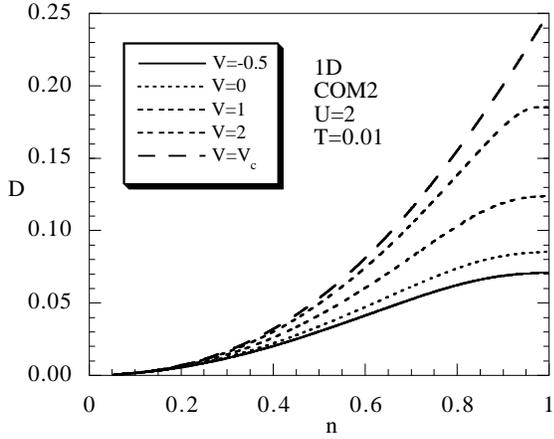}
\end{center}
\caption{The double occupancy versus the particle density at
$T=0.01$, $U=2$ and several values of $V$.}\label{Fig.5.4.1}
\end{figure}

\begin{figure}[tbp]
\begin{center}
\includegraphics[width=7.5cm,clip=]{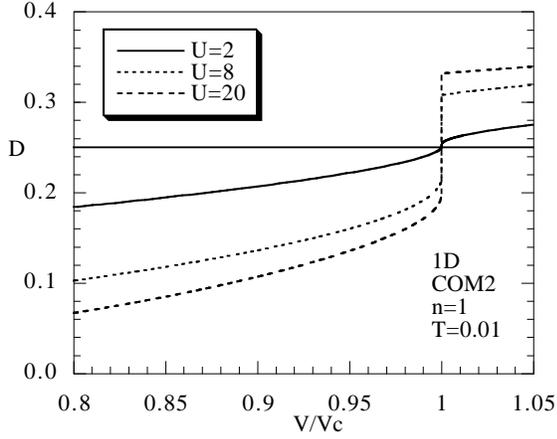}
\end{center}
\caption{The double occupancy at half filling versus $V/V_c$ at
$T=0.01$ and several values of the onsite Coulomb potential.}
\label{Fig.5.4.2} \end{figure}

\begin{figure}[tbp]
\begin{center}
\includegraphics[width=7.5cm,clip=]{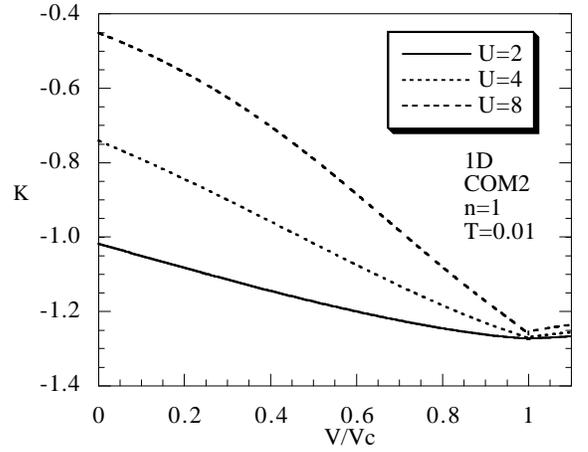}
\hfill
\includegraphics[width=7.5cm,clip=]{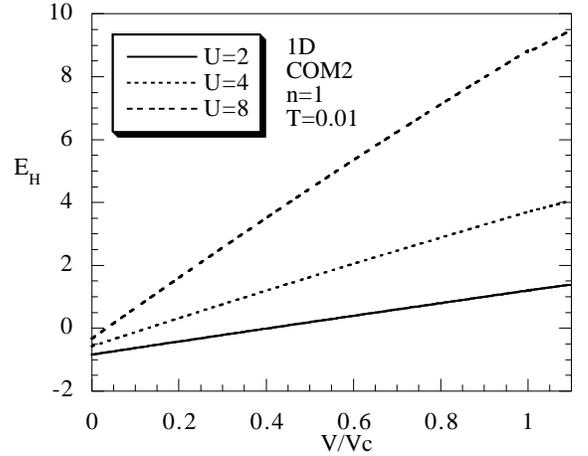}
\end{center}
\caption{The kinetic energy $K$ and the internal energy $E$ are
plotted versus the intersite Coulomb potential, for $n=1$,
$T=0.01$ and various values of the onsite Coulomb potential.}
\label{Fig.5.4.3}
\end{figure}

In Fig.~\ref{Fig.5.4.1} the double occupancy $D$ is shown as a
function of the particle density for $U=2$ and several values of
the intersite Coulomb potential up to the critical value $V_c$
[for $U=2$ we have $V_c\approx 2.47$]. The observed features agree
with the expectation that the double occupancy increases with the
intersite potential. At half filling the double occupancy exhibits
a discontinuity at the critical value $V_c$ [for $U=4$ we have
$V_c\approx 4.978$] (see Fig.~\ref{Fig.5.4.2}). Again, this is a
signal of a phase transition from the Mott insulator to a CO
state. It is interesting to observe that at the middle point of
the jump, the double occupancy takes the value $0.25$ (the
noninteracting one) for all values of $U$. This result can also be
inferred from Fig.~1 in Ref. \onlinecite{Jeckelmann:02}. It seems
as the effect of $V$ completely neutralizes the effect of $U$ and
the system, at least for some local quantities, behaves as the
noninteracting one.
%

From the Hamiltonian (\ref{Eq.5.1.1}) we obtain the internal
energy per site $E_H$
\begin{equation}\label{Eq.5.4.1}
E_H={1 \over N}\langle H \rangle =4dt \langle c^\alpha
(i)c^\dagger (i)\rangle +UD+dV \langle n^\alpha (i)n(i) \rangle
\end{equation}
By increasing the intersite Coulomb potential the kinetic energy
$K=4dt \langle c^\alpha (i)c^\dagger (i)\rangle$ decreases and the
internal energy $E_H$ increases, respectively, owing to the fact
that the double occupancy increases. This is shown in
Fig.~\ref{Fig.5.4.3}. We can clearly see that the kinetic energy,
as a function of the intersite Coulomb potential, shows a minimum
at $V_c$ when the transition is of the second order and develops a
cusp when the transition is of the first order. The internal
energy, instead, is almost insensible to the transition in the
first case, but also develops a small cusp when the transition
becomes of the first order (not shown).

\subsection{Two-dimensional system}

\begin{figure}[tbp]
\begin{center}
\includegraphics[width=7.5cm,clip=]{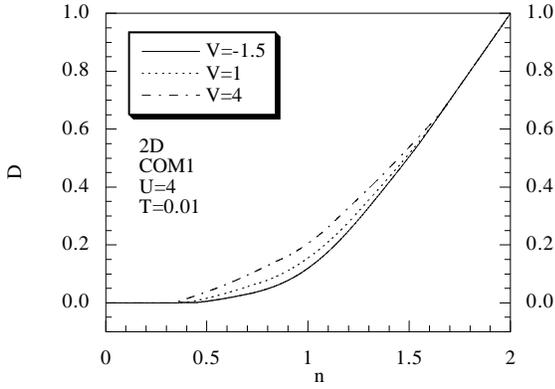}
\end{center}
\caption{The double occupancy versus the particle density at
$T=0.01$, $U=4$ and several values of $V$.} \label{Fig.5.4.4}
\end{figure}

\begin{figure}[tbp]
\begin{center}
\includegraphics[width=7.5cm,clip=]{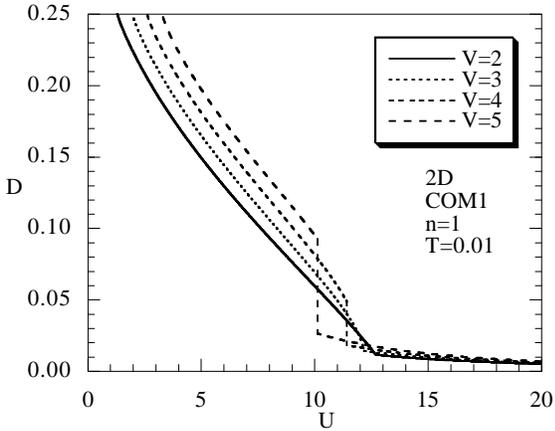}
\hfill
\includegraphics[width=7.5cm,clip=]{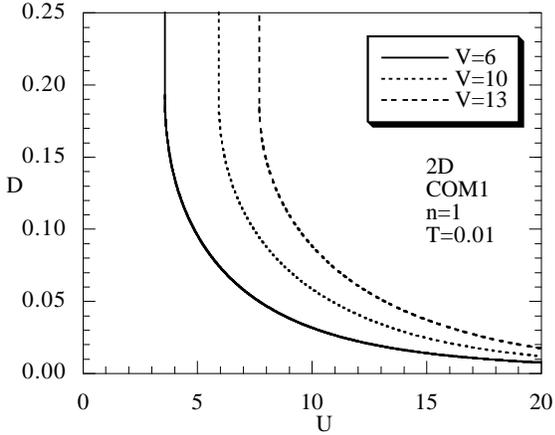}
\end{center}
\caption{The double occupancy versus the onsite potential at
$T=0.01$, half filling and several values of $V$.}
\label{Fig.5.4.5}
\end{figure}

In Fig.~\ref{Fig.5.4.4}, the double occupancy is plotted versus
the filling for different values of the intersite Coulomb
potential. The behavior is very similar to what has been found in
the 1D case. On increasing the intersite Coulomb potential the
double occupancy behaves as that of a non-interacting system
(i.e., tends to $n^2/4$).

The double occupancy as a function of the onsite Coulomb potential
is studied in Figs.~\ref{Fig.5.4.5} (top) and \ref{Fig.5.4.5}
(bottom). In Fig.~\ref{Fig.5.4.5} (top) we observe, by decreasing
the onsite Coulomb potential, the transition from the insulating
phase to the metallic phase and to the charge ordered phase. In
the insulating state the double occupancy is quite depressed; at
the MIT there is a discontinuity and the double occupancy rapidly
increases by decreasing the onsite Coulomb potential until
encounters a second discontinuity when a CO phase comes into play.
The case where $V\ge 6$ is studied in Fig.~\ref{Fig.5.4.5}
(bottom); in accordance with the phase diagram shown in
Fig.~\ref{Fig.5.3.2}, there is no metallic phase; by decreasing
the onsite Coulomb potential, the double occupancy regularly
increases until the charge ordered phase is reached.

\begin{figure}[tbp]
\begin{center}
\includegraphics[width=7.5cm,clip=]{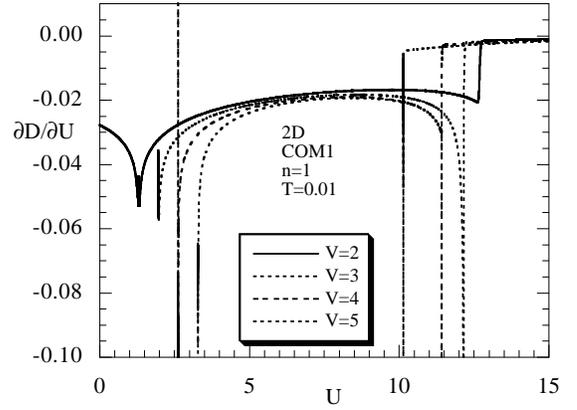}
\hfill
\includegraphics[width=7.5cm,clip=]{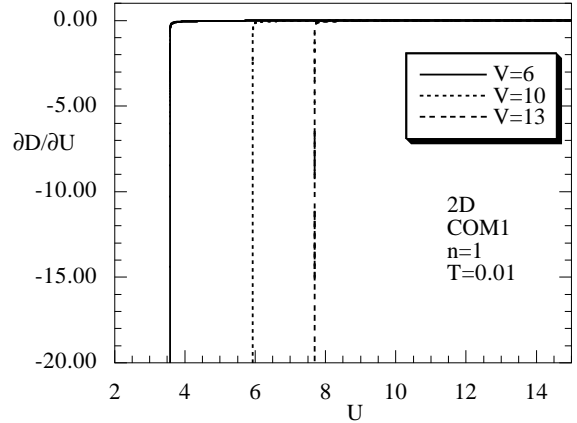}
\end{center}
\caption{The derivative of the double occupancy with respect to
the onsite Coulomb potential is plotted versus the onsite
potential at $T=0.01$, half filling and several values of $V$.}
\label{Fig.5.4.6}
\end{figure}

\begin{figure}[tbp]
\begin{center}
\includegraphics[width=7.5cm,clip=]{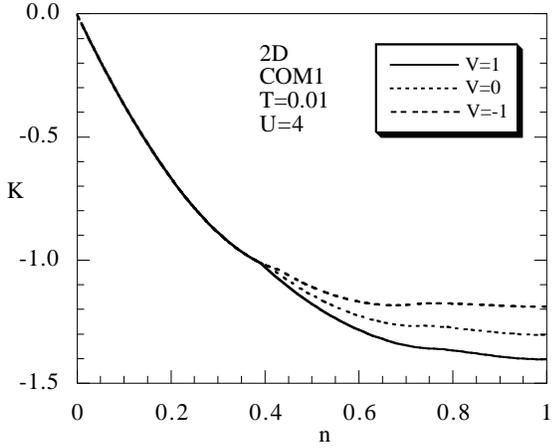}
\end{center}
\caption{The kinetic energy as a function of the particle density
for $U=4$, $T=0.01$ and various values of $V$.} \label{Fig.5.4.8}
\end{figure}

The phase transitions are marked by a discontinuity in the double
occupancy. This is shown in Fig.~\ref{Fig.5.4.6} where $(\partial
D/\partial U)_{n=1}$ is plotted versus the onsite Coulomb
potential for different values of the intersite Coulomb potential.
For $V<6$ (see Fig.~\ref{Fig.5.4.6} (top)) we have two
discontinuities, related to the MIT and insulator to charge
ordered phase transitions. For $V\ge 6$ (see Fig.~\ref{Fig.5.4.6}
(bottom)) there is no metallic state and we observe only one
discontinuity, related to the latter type of transition.

%

\begin{figure}[tbp]
\begin{center}
\includegraphics[width=7.5cm,clip=]{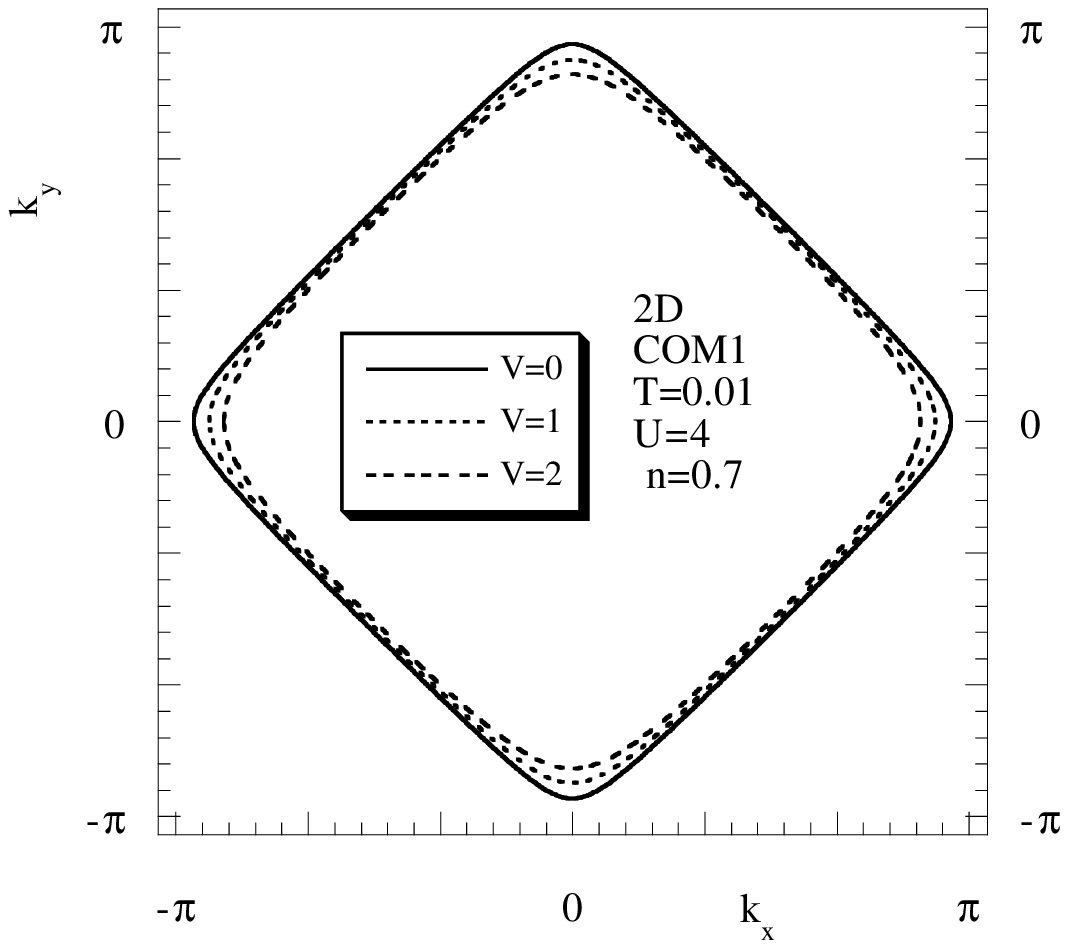}
\hfill
\includegraphics[width=7.5cm,clip=]{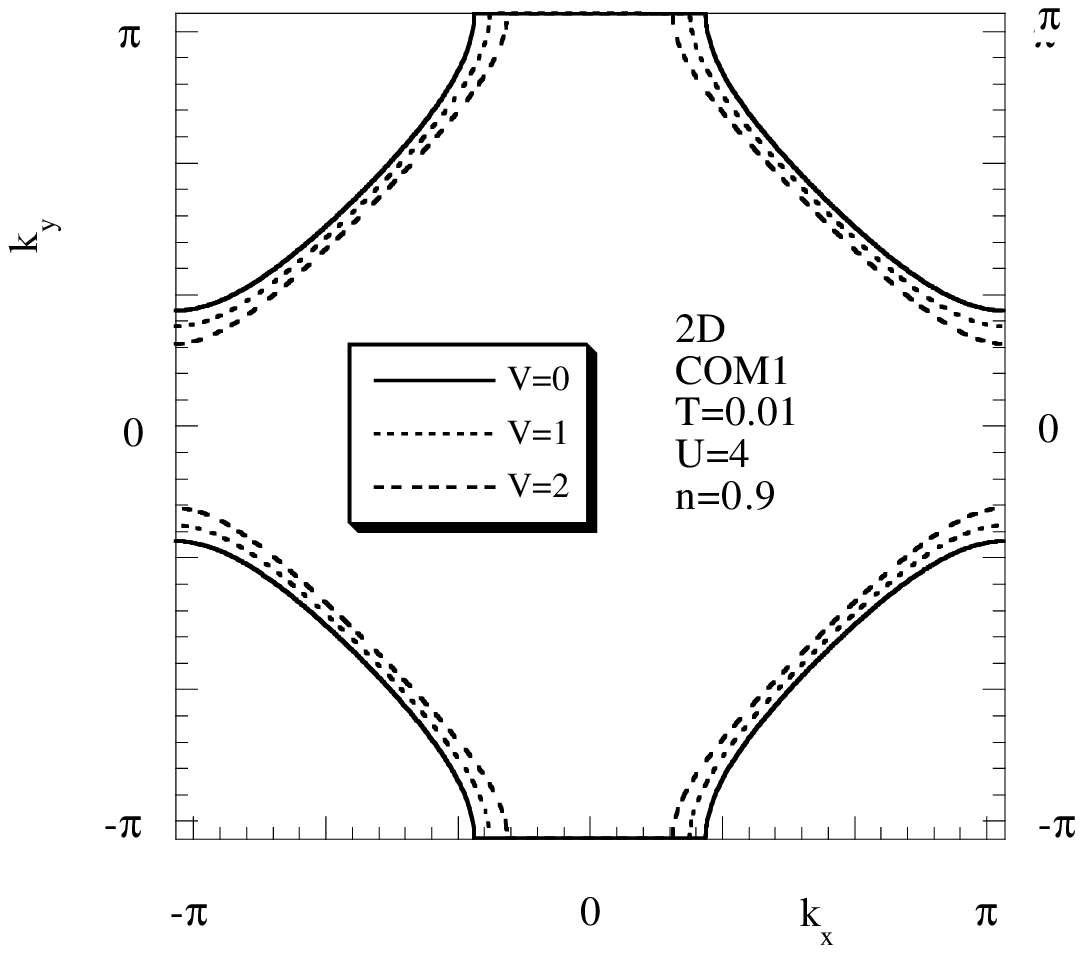}
\end{center}
\caption{The Fermi surface for various values of $V$ at $U=4$ and
$n=0.7$, $0.9$.} \label{Fig.5.5.1}
\end{figure}

\begin{figure}[tbp]
\begin{center}
\includegraphics[width=7.5cm,clip=]{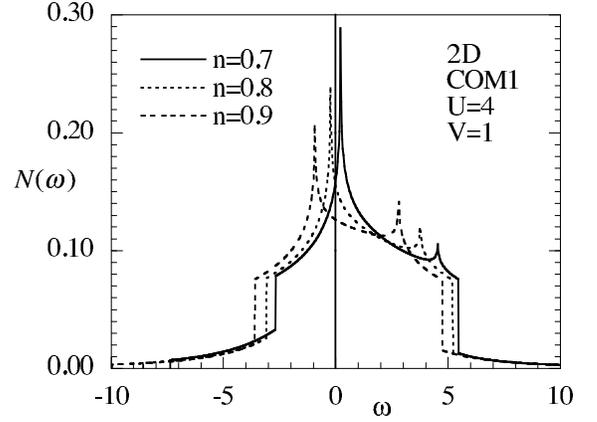}
\end{center}
\caption{The density of states at $U=4$ and $n=0.7$, $0.9$ for
$V=1$.} \label{Fig.5.5.2}
\end{figure}

Accordingly to the behavior of the double occupancy, for positive
values of the intersite Coulomb potential, the mobility of the
electrons increases, signalling an overall tendency towards a
charge density instability driven by the $V$ term (see
Fig.~\ref{Fig.5.4.8}), in agreement with the mean-field result
\cite{Chattopadhyay:97}.

\section{The Fermi Surface and the DOS}

With respect to what we have found in the simple Hubbard
model\cite{Avella:98c}, the overall shape and bending of the Fermi
surface of the system does not change on varying the intersite
Coulomb potential, but, almost rigidly, its volume decreases on
increasing the intersite Coulomb potential [see
Figs.~\ref{Fig.5.5.1}]. This can be explained as an isotropic
increment of the available states in ${\bf k}$-space and can be
useful to describe quantitatively rather than
qualitatively\cite{Avella:98c} (as the usual Hubbard model does)
the ARPES determinations of the Fermi surface of the cuprate
superconductors \cite{Markiewicz:96}.

As we can see from Figs.~\ref{Fig.5.5.1} (top) and \ref{Fig.5.5.1}
(bottom), for given values of the onsite Coulomb potential and the
intersite Coulomb potential, there is a critical value of the
doping $n_c$ where the Fermi surface is nested. At this critical
value there is a crossing of the van Hove singularity and the
Fermi level in the density of states (see Fig.~\ref{Fig.5.5.2}).

\begin{figure}[tbp]
\begin{center}
\includegraphics[width=7.5cm,clip=]{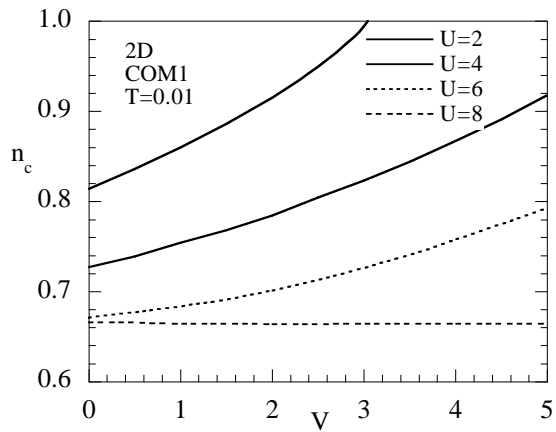}
\end{center}
\caption{The critical value $n_c$ of the doping is shown as a
function of V for various values of U.} \label{Fig.5.5.3}
\end{figure}

This feature may be related, in the framework of the van Hove
scenario, at the experimental results for the static
susceptibility \cite{Torrance:89,Johnston:89} and specific heat
\cite{Loram:89,Wada:90,Loram:93,Loram:94} in cuprate
superconductors which exhibit well-definite peaks as functions of
the filling. We have studied the value of the filling $n_c$, the
result is shown in Fig.~\ref{Fig.5.5.3}, where $n_c$ is plotted as
a function of the intersite Coulomb potential for various values
of the onsite Coulomb potential. As we can see, $n_c$ increases on
increasing the intersite Coulomb potential until a certain value
of $U \approx 8$ is reached. Above this critical value of the
onsite Coulomb potential the influence of the intersite Coulomb
potential is almost zero. This is a clear indication that if we
would like to explain the cuprate superconductors and their
anomalous features by means of Hubbard-like models, we need to
exploit the intermediate regime for the onsite coupling and the
weak regime for the intersite one. Only in this region of the
model parameters, $n_c$ is in qualitative agreement with the
experimental data
\cite{Torrance:89,Johnston:89,Loram:89,Wada:90,Loram:93,Loram:94}.

\section{The Specific Heat}

The specific heat is shown in Fig.~\ref{Fig.5.6.1} for the
one-dimensional case and in Fig.\ref{Fig.5.6.3} for the
two-dimensional one.

\subsection{One-dimensional case}

\begin{figure}[tbp]
\begin{center}
\includegraphics[width=7.5cm,clip=]{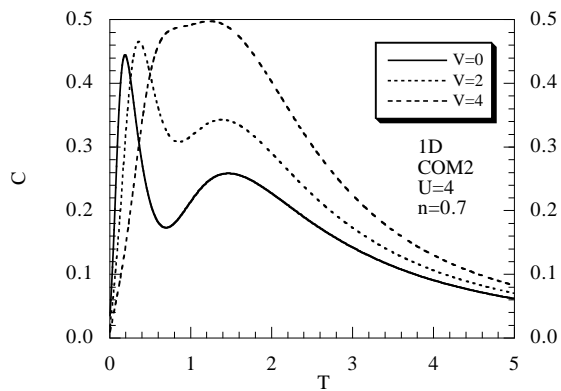}
\end{center}
\caption{The specific heat versus the temperature at $U=4$,
$n=0.7$ and several values of $V$.} \label{Fig.5.6.1}
\end{figure}

\begin{figure}[tbp]
\begin{center}
\includegraphics[width=7.5cm,clip=]{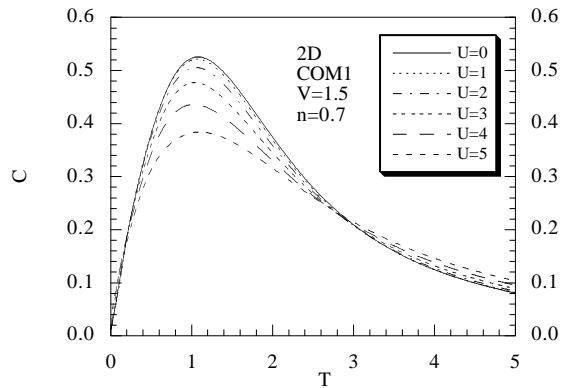}
\end{center}
\caption{The specific heat versus the temperature at $n=0.7$,
$V=1.5$ and several values of $U$.} \label{Fig.5.6.3}
\end{figure}

In 1D, the net effect of the intersite Coulomb repulsion is to
reduce the splitting between the charge and spin energy scales
with respect to what observed for the simple Hubbard model, with a
resulting more pronounced single peak at intermediate temperatures
(see Fig.\ref{Fig.5.6.1}).

\subsection{Two-dimensional case}


As regards the 2D case, it is worth noting that there is a
crossing point when the curves are plotted for different values of
the onsite Coulomb potential at fixed $V$ (see
Fig.\ref{Fig.5.6.3}), but there is no crossing point when the
onsite Coulomb potential is fixed and different values of the
intersite Coulomb potential are considered. This can be seen as an
indication of the fact that the two interaction terms act to
different orders in a perturbation expansion for the double
occupancy and the kinetic energy. In particular, there is no value
of temperature for which the first derivative of the double
occupancy and the kinetic energy with respect to the intersite
Coulomb potential does not depend on the intersite Coulomb
potential.

\section{Conclusions}

The extended Hubbard model, in one and two dimensions, has been
studied by means of the Composite Operator Method within the
2-pole approximation. According to the sign of the intersite
Coulomb potential, transitions between the paramagnetic (metal and
insulator) phase towards two kinds of charge ordered states
(one-particle-per-site and checkerboard types) have been found.
The rank of these transitions has been studied through the
analysis of double occupancy, nearest-neighbor density-density
correlator, kinetic and internal energies behavior. The evolution
of the Fermi surface and of the nesting filling has been tracked
on changing the value of the intersite potential evidencing the
range of parameters suitable to describe the experimental features
of the cuprate superconductors. The specific heat features on
varying the intensity of the intersite potential has been
contrasted to that found for the simple Hubbard model.

\appendix
\section{Spectral density functions and energy spectra}

Within the 2-pole approximation, the energy spectra $E_i({\bf k})$
and the spectral density functions $\sigma ^{(i)}({\bf k})$,
appearing in Eq.~\ref{2pole}, are given by
\begin{align}\label{Eq.5.1.5}
\begin{split}
&E_1({\bf k})=R({\bf k})+Q({\bf k})\\
&E_2({\bf k})=R({\bf k})-Q({\bf k})
\end{split}
\end{align}
\begin{align}
\begin{split}
&\sigma _{11}^{(1)}({\bf k})={{I_{11}} \over 2}\left[ {1+{{g({\bf
k})} \over {2Q({\bf k})}}} \right]\\
&\sigma _{11}^{(2)}({\bf k})={{I_{11}} \over 2}\left[ {1-{{g({\bf
k})} \over {2Q({\bf k})}}} \right]\\
&\sigma _{12}^{(1)}({\bf k})={{m_{12}({\bf k})} \over {2Q({\bf
k})}}
\end{split}
\end{align}
\begin{align}
\begin{split}
&\sigma _{12}^{(2)}({\bf k})=-{{m_{12}({\bf k})} \over {2Q({\bf
k})}}\\
&\sigma _{22}^{(1)}({\bf k})={{I_{22}} \over 2}\left[ {1-{{g({\bf
k})} \over {2Q({\bf k})}}} \right]\\
&\sigma _{22}^{(2)}({\bf k})={{I_{22}} \over 2}\left[ {1+{{g({\bf
k})} \over {2Q({\bf k})}}} \right]
\end{split}
\end{align}
with the notation
\begin{align}\label{Eq.5.1.7}
\begin{split}
&R=R_0+R_1\alpha({\bf k})\\
&R_0=-\mu +{U \over 2}+{d \over {I_{11}I_{22}}} \{ {-t\Delta +{1
\over 2}V[n^2+(1-n)\chi_c^\alpha ] } \}\\
&R_1={d \over {I_{11}I_{22}}}\{
-t[p+I_{22}(1-n)]+V[I_{11}C_{22}^\alpha +I_{22}C_{11}^\alpha ]\}
\end{split}
\end{align}
\begin{align}
\begin{split}
&g({\bf k})=g_0+g_1\alpha ({\bf k})\\ &g_0=-U+{{2d} \over
{I_{11}I_{22}}}[t(1-n)\Delta +{1 \over 2}V(n^2-\chi_c^\alpha )]\\
&g_1={{2d} \over
{I_{11}I_{22}}}[t(1-n)(p-I_{22})+V(I_{22}C_{11}^\alpha
-I_{11}C_{22}^\alpha )]
\end{split}
\end{align}
\begin{align}
\begin{split}
&m_{12}({\bf k})=m_0+m_1\alpha ({\bf k})\\
&m_0=2dt\Delta\\
&m_1=2d[t(p-I_{22})+V C_{12}^\alpha ]\\
&Q({\bf k})={1 \over 2}\sqrt {g^2({\bf k})+{{4m_{12}^2({\bf k})}
\over {I_{11}I_{22}}}}
\end{split}
\end{align}
$I_{11}=1-n/2$ and $I_{22}=n/2$ are the only non-zero entries of
the normalization matrix; the parameters appearing in the previous
equations are defined as
\begin{align}\label{Eq.5.1.10}
\begin{split}
&C_{\mu \nu }^\alpha =\langle \psi _\mu ^\alpha (i)\psi_\nu ^\dagger (i)\rangle\\
&\Delta = C_{11}^\alpha -C_{22}^\alpha
\end{split}
\end{align}
\begin{align}
\begin{split}
&\chi _c^\alpha = \langle n(i)n^\alpha (i) \rangle\\
&p = {1 \over 4} \langle n_\mu ^\alpha (i)n_\mu (i)\rangle -
\langle [c_\uparrow (i)c_\downarrow (i)]^\alpha c_\downarrow
^\dagger (i)c_\uparrow ^\dagger (i) \rangle
\end{split}
\end{align}
We can see that the Green's function depends on the following set
of parameters: $\mu$ , $C_{11}^\alpha$, $C_{12}^\alpha$,
$C_{22}^\alpha$ , $p$, $\chi_c^\alpha$. They can be computed as
functions of the model parameters and temperature and filling by
the following set of coupled self-consistent equations
\begin{align}\label{Eq.5.1.11}
\begin{split}
&n=2[1-C_{11}-C_{22}]\\
&C_{\mu \nu }^\alpha =\langle \psi_\mu^\alpha (i)\psi_\nu^\dagger (i) \rangle
\end{split}
\end{align}
\begin{align}\label{Pauli}
\begin{split}
&C_{12}=0\\
&\chi_c^\alpha =n^2-{{n(2-n)} \over
{n-2D}}I_{11}^{-1}(C_{11}^\alpha +C_{12}^\alpha )^2 \\
&-{{n(2-n)} \over {n-2D}} I_{22}^{-1}(C_{12}^\alpha +C_{22}^\alpha
)^2
\end{split}
\end{align}
where $C_{\mu \nu }= \langle \psi _\mu (i)\psi _\nu ^\dagger (i)
\rangle$ and $D=\frac1N \sum_{\bf i} \langle n_\uparrow({\bf i})
n_\downarrow({\bf i}) \rangle = \frac n2- C_{22}$ is the double
occupancy.

The correlation function $C(i,j)=\langle \psi (i)\psi ^\dagger (j)
\rangle$ can be computed in terms of the retarded propagator, or
better in terms of the energy spectra and of the spectral density
functions, by means of the following expression \cite{Mancini:00}
\begin{multline}\label{Eq.5.1.12}
C(i,j)={\Omega \over {2(2\pi )^d}}\sum_{n=1}^2 \int_{\Omega_B}
d^dk\,e^{{\rm i}[{\bf k} \cdot ({\bf i}-{\bf j})-E_n({\bf
k})(t_i-t_j)]} \times \\
\times \sigma ^{(n)}({\bf k})[1+T_n({\bf k})]
\end{multline}
where $\Omega$ and $\Omega_B$ are the volume of the unit cells in
the direct and inverse spaces, respectively, and $T_n({\bf
k})=\tanh \left( {E_n({\bf k})/2k_BT} \right)$.


\begin{thebibliography}{71}
\expandafter\ifx\csname
natexlab\endcsname\relax\def\natexlab#1{#1}\fi
\expandafter\ifx\csname bibnamefont\endcsname\relax
  \def\bibnamefont#1{#1}\fi
\expandafter\ifx\csname bibfnamefont\endcsname\relax
  \def\bibfnamefont#1{#1}\fi
\expandafter\ifx\csname citenamefont\endcsname\relax
  \def\citenamefont#1{#1}\fi
\expandafter\ifx\csname url\endcsname\relax
  \def\url#1{\texttt{#1}}\fi
\expandafter\ifx\csname
urlprefix\endcsname\relax\def\urlprefix{URL }\fi
\providecommand{\bibinfo}[2]{#2}
\providecommand{\eprint}[2][]{\url{#2}}

\bibitem[{\citenamefont{Varma}(1987)}]{Varma:87}
\bibinfo{author}{\bibfnamefont{C.}~\bibnamefont{Varma}},
  \bibinfo{journal}{Sol.~Stat.~Comm.} \textbf{\bibinfo{volume}{62}},
  \bibinfo{pages}{681} (\bibinfo{year}{1987}).

\bibitem[{\citenamefont{Littlewood et~al.}(1989)\citenamefont{Littlewood,
  Varma, and Abrahams}}]{Littlewood:89}
\bibinfo{author}{\bibfnamefont{P.}~\bibnamefont{Littlewood}},
  \bibinfo{author}{\bibfnamefont{C.}~\bibnamefont{Varma}}, \bibnamefont{and}
  \bibinfo{author}{\bibfnamefont{E.}~\bibnamefont{Abrahams}},
  \bibinfo{journal}{Phys.~Rev.~Lett.} \textbf{\bibinfo{volume}{63}},
  \bibinfo{pages}{2602} (\bibinfo{year}{1989}).

\bibitem[{\citenamefont{Varma}(1995)}]{Varma:95}
\bibinfo{author}{\bibfnamefont{C.}~\bibnamefont{Varma}},
  \bibinfo{journal}{Phys.~Rev.~Lett.} \textbf{\bibinfo{volume}{75}},
  \bibinfo{pages}{898} (\bibinfo{year}{1995}).

\bibitem[{\citenamefont{Janner}(1995)}]{Janner:95}
\bibinfo{author}{\bibfnamefont{A.}~\bibnamefont{Janner}},
  \bibinfo{journal}{Phys.~Rev.~B} \textbf{\bibinfo{volume}{52}},
  \bibinfo{pages}{17158} (\bibinfo{year}{1995}).

\bibitem[{\citenamefont{{van den Brink} et~al.}(1995)}]{vandenBrink:95}
\bibinfo{author}{\bibfnamefont{J.}~\bibnamefont{{van den Brink}}}
  \bibnamefont{et~al.}, \bibinfo{journal}{Phys.~Rev.~Lett.}
  \textbf{\bibinfo{volume}{75}}, \bibinfo{pages}{4658} (\bibinfo{year}{1995}).

\bibitem[{Hub()}]{Hubbard}
\bibinfo{note}{J.~Hubbard, Proc.~Roy.~Soc.~A \textbf{276}, 238 (1963);
  \textbf{277}, 237 (1964); \textbf{281}, 401 (1964); \textbf{285}, 542
  (1965).}

\bibitem[{\citenamefont{{van den Brink} et~al.}(1997)\citenamefont{{van den
  Brink}, Eder, and Sawatzky}}]{vandenBrink:97}
\bibinfo{author}{\bibfnamefont{J.}~\bibnamefont{{van den Brink}}},
  \bibinfo{author}{\bibfnamefont{R.}~\bibnamefont{Eder}}, \bibnamefont{and}
  \bibinfo{author}{\bibfnamefont{G.}~\bibnamefont{Sawatzky}},
  \bibinfo{journal}{Europhys.~Lett.} \textbf{\bibinfo{volume}{37}},
  \bibinfo{pages}{471} (\bibinfo{year}{1997}).

\bibitem[{\citenamefont{Sim{\'o}n et~al.}(1997)\citenamefont{Sim{\'o}n, Aligia,
  and Gagliano}}]{Simon:97}
\bibinfo{author}{\bibfnamefont{M.}~\bibnamefont{Sim{\'o}n}},
  \bibinfo{author}{\bibfnamefont{A.}~\bibnamefont{Aligia}}, \bibnamefont{and}
  \bibinfo{author}{\bibfnamefont{E.}~\bibnamefont{Gagliano}},
  \bibinfo{journal}{Phys.~Rev.~B} \textbf{\bibinfo{volume}{56}},
  \bibinfo{pages}{5637} (\bibinfo{year}{1997}).

\bibitem[{\citenamefont{Ferrer et~al.}(1998)\citenamefont{Ferrer,
  Gonz{\'a}les-Alvarez, and Sanchez-Canizares}}]{Ferrer:98}
\bibinfo{author}{\bibfnamefont{J.}~\bibnamefont{Ferrer}},
  \bibinfo{author}{\bibfnamefont{M.}~\bibnamefont{Gonz{\'a}les-Alvarez}},
  \bibnamefont{and}
  \bibinfo{author}{\bibfnamefont{J.}~\bibnamefont{Sanchez-Canizares}},
  \bibinfo{journal}{Phys.~Rev.~B} \textbf{\bibinfo{volume}{57}},
  \bibinfo{pages}{7470} (\bibinfo{year}{1998}).

\bibitem[{\citenamefont{Wigner}(1938)}]{Wigner:38}
\bibinfo{author}{\bibfnamefont{E.}~\bibnamefont{Wigner}},
  \bibinfo{journal}{Trans.~Faraday~Soc.} \textbf{\bibinfo{volume}{34}},
  \bibinfo{pages}{678} (\bibinfo{year}{1938}).

\bibitem[{\citenamefont{Fulde}(1997)}]{Fulde:97}
\bibinfo{author}{\bibfnamefont{P.}~\bibnamefont{Fulde}},
  \bibinfo{journal}{Ann.~Phys.} \textbf{\bibinfo{volume}{6}},
  \bibinfo{pages}{178} (\bibinfo{year}{1997}).

\bibitem[{\citenamefont{Andrei et~al.}(1988)}]{Andrei:88}
\bibinfo{author}{\bibfnamefont{E.~Y.} \bibnamefont{Andrei}}
  \bibnamefont{et~al.}, \bibinfo{journal}{Phys.~Rev.~Lett.}
  \textbf{\bibinfo{volume}{60}}, \bibinfo{pages}{2765} (\bibinfo{year}{1988}).

\bibitem[{\citenamefont{Ochiai et~al.}(1990)\citenamefont{Ochiai, Suzuki, and
  Kasuya}}]{Ochiai:90}
\bibinfo{author}{\bibfnamefont{A.}~\bibnamefont{Ochiai}},
  \bibinfo{author}{\bibfnamefont{T.}~\bibnamefont{Suzuki}}, \bibnamefont{and}
  \bibinfo{author}{\bibfnamefont{T.}~\bibnamefont{Kasuya}},
  \bibinfo{journal}{J.~Phys.~Soc.~Jpn.} \textbf{\bibinfo{volume}{59}},
  \bibinfo{pages}{4129} (\bibinfo{year}{1990}).

\bibitem[{\citenamefont{Chen and Cheong}(1996)}]{Chen:96}
\bibinfo{author}{\bibfnamefont{C.~H.} \bibnamefont{Chen}} \bibnamefont{and}
  \bibinfo{author}{\bibfnamefont{S.~W.} \bibnamefont{Cheong}},
  \bibinfo{journal}{Phys.~Rev.~Lett.} \textbf{\bibinfo{volume}{76}},
  \bibinfo{pages}{4042} (\bibinfo{year}{1996}).

\bibitem[{\citenamefont{Ohama et~al.}(1999)}]{Ohama:99}
\bibinfo{author}{\bibfnamefont{T.}~\bibnamefont{Ohama}} \bibnamefont{et~al.},
  \bibinfo{journal}{Phys.~Rev.~B} \textbf{\bibinfo{volume}{59}},
  \bibinfo{pages}{3299} (\bibinfo{year}{1999}).

\bibitem[{\citenamefont{Vojta et~al.}(2000)}]{Vojta:00}
\bibinfo{author}{\bibfnamefont{M.}~\bibnamefont{Vojta}} \bibnamefont{et~al.},
  \bibinfo{journal}{Phys.~Rev.~B} \textbf{\bibinfo{volume}{62}},
  \bibinfo{pages}{6721} (\bibinfo{year}{2000}).

\bibitem[{\citenamefont{Salamon et~al.}(2001)}]{Salamon:01}
\bibinfo{author}{\bibfnamefont{M.}~\bibnamefont{Salamon}} \bibnamefont{et~al.},
  \bibinfo{journal}{Rev.~Mod.~Phys.} \textbf{\bibinfo{volume}{73}},
  \bibinfo{pages}{583} (\bibinfo{year}{2001}).

\bibitem[{\citenamefont{Park et~al.}(1998)}]{Park:98}
\bibinfo{author}{\bibfnamefont{S.~K.} \bibnamefont{Park}} \bibnamefont{et~al.},
  \bibinfo{journal}{Phys.~Rev.~B} \textbf{\bibinfo{volume}{58}},
  \bibinfo{pages}{3717} (\bibinfo{year}{1998}).

\bibitem[{\citenamefont{Ueda et~al.}(2001)}]{Ueda:01}
\bibinfo{author}{\bibfnamefont{Y.}~\bibnamefont{Ueda}} \bibnamefont{et~al.},
  \bibinfo{journal}{J.~All.~Comp.} \textbf{\bibinfo{volume}{317}},
  \bibinfo{pages}{109} (\bibinfo{year}{2001}).

\bibitem[{\citenamefont{Chow et~al.}(2000)}]{Chow:00}
\bibinfo{author}{\bibfnamefont{D.~S.} \bibnamefont{Chow}} \bibnamefont{et~al.},
  \bibinfo{journal}{Phys.~Rev.~Lett.} \textbf{\bibinfo{volume}{85}},
  \bibinfo{pages}{1698} (\bibinfo{year}{2000}).

\bibitem[{\citenamefont{Seo and Fukuyama}(1998)}]{Seo:98}
\bibinfo{author}{\bibfnamefont{H.}~\bibnamefont{Seo}} \bibnamefont{and}
  \bibinfo{author}{\bibfnamefont{H.}~\bibnamefont{Fukuyama}},
  \bibinfo{journal}{J.~Phys.~Soc.~Jpn.} \textbf{\bibinfo{volume}{67}},
  \bibinfo{pages}{2602} (\bibinfo{year}{1998}).

\bibitem[{\citenamefont{Dongen}(1994)}]{vanDongen:94a}
\bibinfo{author}{\bibfnamefont{P.~V.} \bibnamefont{Dongen}},
  \bibinfo{journal}{Phys.~Rev.~B} \textbf{\bibinfo{volume}{50}},
  \bibinfo{pages}{14016} (\bibinfo{year}{1994}).

\bibitem[{\citenamefont{Pietig et~al.}(1999)\citenamefont{Pietig, Bulla, and
  Blawid}}]{Pietig:99}
\bibinfo{author}{\bibfnamefont{R.}~\bibnamefont{Pietig}},
  \bibinfo{author}{\bibfnamefont{R.}~\bibnamefont{Bulla}}, \bibnamefont{and}
  \bibinfo{author}{\bibfnamefont{S.}~\bibnamefont{Blawid}},
  \bibinfo{journal}{Phys.~Rev.~Lett.} \textbf{\bibinfo{volume}{82}},
  \bibinfo{pages}{4046} (\bibinfo{year}{1999}).

\bibitem[{\citenamefont{McKenzie et~al.}(2001)}]{McKenzie:01}
\bibinfo{author}{\bibfnamefont{R.~H.} \bibnamefont{McKenzie}}
  \bibnamefont{et~al.}, \bibinfo{journal}{Phys.~Rev.~B}
  \textbf{\bibinfo{volume}{64}}, \bibinfo{pages}{085109}
  (\bibinfo{year}{2001}).

\bibitem[{\citenamefont{Merino and McKenzie}(2001)}]{Merino:01}
\bibinfo{author}{\bibfnamefont{J.}~\bibnamefont{Merino}} \bibnamefont{and}
  \bibinfo{author}{\bibfnamefont{R.~H.} \bibnamefont{McKenzie}},
  \bibinfo{journal}{Phys.~Rev.~Lett.} \textbf{\bibinfo{volume}{87}},
  \bibinfo{pages}{237002} (\bibinfo{year}{2001}).

\bibitem[{\citenamefont{Hoang and Thalmeier}(2002)}]{Hoang:02}
\bibinfo{author}{\bibfnamefont{A.}~\bibnamefont{Hoang}} \bibnamefont{and}
  \bibinfo{author}{\bibfnamefont{P.}~\bibnamefont{Thalmeier}},
  \bibinfo{journal}{J.~Phys.:~Condens.~Matter} \textbf{\bibinfo{volume}{14}},
  \bibinfo{pages}{6639} (\bibinfo{year}{2002}).

\bibitem[{\citenamefont{Hirsch}(1984)}]{Hirsch:84a}
\bibinfo{author}{\bibfnamefont{J.}~\bibnamefont{Hirsch}},
  \bibinfo{journal}{Phys.~Rev.~Lett.} \textbf{\bibinfo{volume}{53}},
  \bibinfo{pages}{2327} (\bibinfo{year}{1984}).

\bibitem[{\citenamefont{Hellberg}(2001)}]{Hellberg:01}
\bibinfo{author}{\bibfnamefont{C.}~\bibnamefont{Hellberg}},
  \bibinfo{journal}{J.~Appl.~Phys.} \textbf{\bibinfo{volume}{89}},
  \bibinfo{pages}{6627} (\bibinfo{year}{2001}).

\bibitem[{\citenamefont{Calandra et~al.}(2002)\citenamefont{Calandra, Merino,
  and McKenzie}}]{Calandra:02}
\bibinfo{author}{\bibfnamefont{M.}~\bibnamefont{Calandra}},
  \bibinfo{author}{\bibfnamefont{J.}~\bibnamefont{Merino}}, \bibnamefont{and}
  \bibinfo{author}{\bibfnamefont{R.~H.} \bibnamefont{McKenzie}},
  \bibinfo{journal}{Phys.~Rev.~B} \textbf{\bibinfo{volume}{66}},
  \bibinfo{pages}{195102} (\bibinfo{year}{2002}).

\bibitem[{\citenamefont{Mancini and Avella}(2003)}]{Mancini:00}
\bibinfo{author}{\bibfnamefont{F.}~\bibnamefont{Mancini}} \bibnamefont{and}
  \bibinfo{author}{\bibfnamefont{A.}~\bibnamefont{Avella}},
  \bibinfo{journal}{Eur.~Phys.~J.~B} \textbf{\bibinfo{volume}{36}},
  \bibinfo{pages}{37} (\bibinfo{year}{2003}).

\bibitem[{\citenamefont{Mancini and Avella}(2004)}]{Mancini:04}
\bibinfo{author}{\bibfnamefont{F.}~\bibnamefont{Mancini}} \bibnamefont{and}
  \bibinfo{author}{\bibfnamefont{A.}~\bibnamefont{Avella}}
  (\bibinfo{year}{2004}), \bibinfo{note}{preprint of the University of
  Salerno}.

\bibitem[{\citenamefont{Tomioka et~al.}(1997)}]{Tomioka:97}
\bibinfo{author}{\bibfnamefont{Y.}~\bibnamefont{Tomioka}} \bibnamefont{et~al.},
  \bibinfo{journal}{J.~Phys.~Soc.~Jpn.} \textbf{\bibinfo{volume}{66}},
  \bibinfo{pages}{302} (\bibinfo{year}{1997}).

\bibitem[{\citenamefont{Chatterji et~al.}(2000)}]{Chatterji:00}
\bibinfo{author}{\bibfnamefont{T.}~\bibnamefont{Chatterji}}
  \bibnamefont{et~al.}, \bibinfo{journal}{Phys.~Rev.~B}
  \textbf{\bibinfo{volume}{61}}, \bibinfo{pages}{570} (\bibinfo{year}{2000}).

\bibitem[{\citenamefont{Dho et~al.}(2001)}]{Dho:01}
\bibinfo{author}{\bibfnamefont{J.}~\bibnamefont{Dho}} \bibnamefont{et~al.},
  \bibinfo{journal}{J. Phys.: Cond. Matt.} \textbf{\bibinfo{volume}{13}},
  \bibinfo{pages}{3655} (\bibinfo{year}{2001}).

\bibitem[{\citenamefont{Hirsch and Scalapino}(1984)}]{Hirsch:84}
\bibinfo{author}{\bibfnamefont{J.~E.} \bibnamefont{Hirsch}} \bibnamefont{and}
  \bibinfo{author}{\bibfnamefont{D.~J.} \bibnamefont{Scalapino}},
  \bibinfo{journal}{Phys.~Rev.~B} \textbf{\bibinfo{volume}{29}},
  \bibinfo{pages}{5554} (\bibinfo{year}{1984}).

\bibitem[{\citenamefont{Emery}(1979)}]{Emery:79}
\bibinfo{author}{\bibfnamefont{V.}~\bibnamefont{Emery}}, in
  \emph{\bibinfo{booktitle}{Highly Conducting One-Dimensional Solids}}, edited
  by \bibinfo{editor}{\bibfnamefont{J.}~\bibnamefont{Devreese}},
  \bibinfo{editor}{\bibfnamefont{R.}~\bibnamefont{Evrand}}, \bibnamefont{and}
  \bibinfo{editor}{\bibfnamefont{V.}~\bibnamefont{{v}an Doren}}
  (\bibinfo{publisher}{Plenum Press}, \bibinfo{address}{New York},
  \bibinfo{year}{1979}), p. \bibinfo{pages}{247}.

\bibitem[{\citenamefont{Emery}(1987)}]{Emery:87}
\bibinfo{author}{\bibfnamefont{V.}~\bibnamefont{Emery}},
  \bibinfo{journal}{Phys.~Rev.~Lett.} \textbf{\bibinfo{volume}{58}},
  \bibinfo{pages}{2794} (\bibinfo{year}{1987}).

\bibitem[{\citenamefont{Zhang and Rice}(1988)}]{Zhang:88}
\bibinfo{author}{\bibfnamefont{F.}~\bibnamefont{Zhang}} \bibnamefont{and}
  \bibinfo{author}{\bibfnamefont{T.}~\bibnamefont{Rice}},
  \bibinfo{journal}{Phys.~Rev.~B} \textbf{\bibinfo{volume}{37}},
  \bibinfo{pages}{3759} (\bibinfo{year}{1988}).

\bibitem[{\citenamefont{Bosch and Falicov}(1988)}]{Bosch:88}
\bibinfo{author}{\bibfnamefont{L.~M.~D.} \bibnamefont{Bosch}} \bibnamefont{and}
  \bibinfo{author}{\bibfnamefont{L.}~\bibnamefont{Falicov}},
  \bibinfo{journal}{Phys.~Rev.~B} \textbf{\bibinfo{volume}{37}},
  \bibinfo{pages}{6073} (\bibinfo{year}{1988}).

\bibitem[{\citenamefont{Zhang and Callaway}(1989)}]{Zhang:89}
\bibinfo{author}{\bibfnamefont{Y.}~\bibnamefont{Zhang}} \bibnamefont{and}
  \bibinfo{author}{\bibfnamefont{J.}~\bibnamefont{Callaway}},
  \bibinfo{journal}{Phys.~Rev.~B} \textbf{\bibinfo{volume}{39}},
  \bibinfo{pages}{9397} (\bibinfo{year}{1989}).

\bibitem[{\citenamefont{Yan}(1993)}]{Yan:93}
\bibinfo{author}{\bibfnamefont{X.-Z.} \bibnamefont{Yan}},
  \bibinfo{journal}{Phys.~Rev.~B} \textbf{\bibinfo{volume}{48}},
  \bibinfo{pages}{7140} (\bibinfo{year}{1993}).

\bibitem[{\citenamefont{Avella et~al.}(2004)\citenamefont{Avella, Krivenko,
  Mancini, and Plakida}}]{Avella:03c}
\bibinfo{author}{\bibfnamefont{A.}~\bibnamefont{Avella}},
  \bibinfo{author}{\bibfnamefont{S.}~\bibnamefont{Krivenko}},
  \bibinfo{author}{\bibfnamefont{F.}~\bibnamefont{Mancini}}, \bibnamefont{and}
  \bibinfo{author}{\bibfnamefont{N.}~\bibnamefont{Plakida}},
  \bibinfo{journal}{J.~Magn.~Magn.~Mat.} \textbf{\bibinfo{volume}{272}},
  \bibinfo{pages}{456} (\bibinfo{year}{2004}).

\bibitem[{\citenamefont{Matsumoto et~al.}(1996)\citenamefont{Matsumoto,
  Saikawa, and Mancini}}]{Matsumoto:96}
\bibinfo{author}{\bibfnamefont{H.}~\bibnamefont{Matsumoto}},
  \bibinfo{author}{\bibfnamefont{T.}~\bibnamefont{Saikawa}}, \bibnamefont{and}
  \bibinfo{author}{\bibfnamefont{F.}~\bibnamefont{Mancini}},
  \bibinfo{journal}{Phys.~Rev.~B} \textbf{\bibinfo{volume}{54}},
  \bibinfo{pages}{14445} (\bibinfo{year}{1996}).

\bibitem[{\citenamefont{Matsumoto and Mancini}(1997)}]{Matsumoto:97}
\bibinfo{author}{\bibfnamefont{H.}~\bibnamefont{Matsumoto}} \bibnamefont{and}
  \bibinfo{author}{\bibfnamefont{F.}~\bibnamefont{Mancini}},
  \bibinfo{journal}{Phys.~Rev.~B} \textbf{\bibinfo{volume}{55}},
  \bibinfo{pages}{2095} (\bibinfo{year}{1997}).

\bibitem[{\citenamefont{Mancini
  et~al.}(1995{\natexlab{a}})\citenamefont{Mancini, Marra, and
  Matsumoto}}]{Mancini:95b}
\bibinfo{author}{\bibfnamefont{F.}~\bibnamefont{Mancini}},
  \bibinfo{author}{\bibfnamefont{S.}~\bibnamefont{Marra}}, \bibnamefont{and}
  \bibinfo{author}{\bibfnamefont{H.}~\bibnamefont{Matsumoto}},
  \bibinfo{journal}{Physica~C} \textbf{\bibinfo{volume}{252}},
  \bibinfo{pages}{361} (\bibinfo{year}{1995}{\natexlab{a}}).

\bibitem[{\citenamefont{Avella et~al.}(1998{\natexlab{a}})\citenamefont{Avella,
  Mancini, S{\'a}nchez-Lopez, Villani, and Buzatu}}]{Avella:98b}
\bibinfo{author}{\bibfnamefont{A.}~\bibnamefont{Avella}},
  \bibinfo{author}{\bibfnamefont{F.}~\bibnamefont{Mancini}},
  \bibinfo{author}{\bibfnamefont{M.}~\bibnamefont{S{\'a}nchez-Lopez}},
  \bibinfo{author}{\bibfnamefont{D.}~\bibnamefont{Villani}}, \bibnamefont{and}
  \bibinfo{author}{\bibfnamefont{F.~D.} \bibnamefont{Buzatu}},
  \bibinfo{journal}{J.~Phys.~Studies} \textbf{\bibinfo{volume}{2}},
  \bibinfo{pages}{228} (\bibinfo{year}{1998}{\natexlab{a}}).

\bibitem[{\citenamefont{S{\'a}nchez-Lopez
  et~al.}(1998)\citenamefont{S{\'a}nchez-Lopez, Avella, and
  Mancini}}]{Avella:98e}
\bibinfo{author}{\bibfnamefont{M.}~\bibnamefont{S{\'a}nchez-Lopez}},
  \bibinfo{author}{\bibfnamefont{A.}~\bibnamefont{Avella}}, \bibnamefont{and}
  \bibinfo{author}{\bibfnamefont{F.}~\bibnamefont{Mancini}},
  \bibinfo{journal}{Europhys.~Lett.} \textbf{\bibinfo{volume}{44}},
  \bibinfo{pages}{328} (\bibinfo{year}{1998}).

\bibitem[{\citenamefont{S{\'a}nchez-Lopez
  et~al.}(1999)\citenamefont{S{\'a}nchez-Lopez, Avella, and
  Mancini}}]{Sanchez:99}
\bibinfo{author}{\bibfnamefont{M.}~\bibnamefont{S{\'a}nchez-Lopez}},
  \bibinfo{author}{\bibfnamefont{A.}~\bibnamefont{Avella}}, \bibnamefont{and}
  \bibinfo{author}{\bibfnamefont{F.}~\bibnamefont{Mancini}},
  \bibinfo{journal}{Physica~B} \textbf{\bibinfo{volume}{259}},
  \bibinfo{pages}{753} (\bibinfo{year}{1999}).

\bibitem[{\citenamefont{Avella et~al.}(2002)\citenamefont{Avella, Mancini, and
  S{\`a}nchez-Lopez}}]{Avella:00}
\bibinfo{author}{\bibfnamefont{A.}~\bibnamefont{Avella}},
  \bibinfo{author}{\bibfnamefont{F.}~\bibnamefont{Mancini}}, \bibnamefont{and}
  \bibinfo{author}{\bibfnamefont{M.}~\bibnamefont{S{\`a}nchez-Lopez}},
  \bibinfo{journal}{Eur.~Phys.~J.~B} \textbf{\bibinfo{volume}{29}},
  \bibinfo{pages}{399} (\bibinfo{year}{2002}).

\bibitem[{\citenamefont{Mancini
  et~al.}(1995{\natexlab{b}})\citenamefont{Mancini, Marra, and
  Matsumoto}}]{Mancini:95}
\bibinfo{author}{\bibfnamefont{F.}~\bibnamefont{Mancini}},
  \bibinfo{author}{\bibfnamefont{S.}~\bibnamefont{Marra}}, \bibnamefont{and}
  \bibinfo{author}{\bibfnamefont{H.}~\bibnamefont{Matsumoto}},
  \bibinfo{journal}{Physica~C} \textbf{\bibinfo{volume}{244}},
  \bibinfo{pages}{49} (\bibinfo{year}{1995}{\natexlab{b}}).

\bibitem[{\citenamefont{Mancini
  et~al.}(1995{\natexlab{c}})\citenamefont{Mancini, Marra, and
  Matsumoto}}]{Mancini:95a}
\bibinfo{author}{\bibfnamefont{F.}~\bibnamefont{Mancini}},
  \bibinfo{author}{\bibfnamefont{S.}~\bibnamefont{Marra}}, \bibnamefont{and}
  \bibinfo{author}{\bibfnamefont{H.}~\bibnamefont{Matsumoto}},
  \bibinfo{journal}{Physica~C} \textbf{\bibinfo{volume}{250}},
  \bibinfo{pages}{184} (\bibinfo{year}{1995}{\natexlab{c}}).

\bibitem[{\citenamefont{Avella et~al.}(1998{\natexlab{b}})\citenamefont{Avella,
  Mancini, Villani, Siurakshina, and Yushankhai}}]{Avella:98}
\bibinfo{author}{\bibfnamefont{A.}~\bibnamefont{Avella}},
  \bibinfo{author}{\bibfnamefont{F.}~\bibnamefont{Mancini}},
  \bibinfo{author}{\bibfnamefont{D.}~\bibnamefont{Villani}},
  \bibinfo{author}{\bibfnamefont{L.}~\bibnamefont{Siurakshina}},
  \bibnamefont{and} \bibinfo{author}{\bibfnamefont{V.~Y.}
  \bibnamefont{Yushankhai}}, \bibinfo{journal}{Int.~J.~Mod.~Phys.~B}
  \textbf{\bibinfo{volume}{12}}, \bibinfo{pages}{81}
  (\bibinfo{year}{1998}{\natexlab{b}}).

\bibitem[{\citenamefont{Avella et~al.}(1998{\natexlab{c}})\citenamefont{Avella,
  Mancini, and S{\'a}nchez-Lopez}}]{Avella:98f}
\bibinfo{author}{\bibfnamefont{A.}~\bibnamefont{Avella}},
  \bibinfo{author}{\bibfnamefont{F.}~\bibnamefont{Mancini}}, \bibnamefont{and}
  \bibinfo{author}{\bibfnamefont{M.}~\bibnamefont{S{\'a}nchez-Lopez}},
  \bibinfo{journal}{J.~Phys.~Studies} \textbf{\bibinfo{volume}{2}},
  \bibinfo{pages}{232} (\bibinfo{year}{1998}{\natexlab{c}}).

\bibitem[{\citenamefont{Mancini et~al.}(1999)\citenamefont{Mancini, Matsumoto,
  and Villani}}]{Mancini:99a}
\bibinfo{author}{\bibfnamefont{F.}~\bibnamefont{Mancini}},
  \bibinfo{author}{\bibfnamefont{H.}~\bibnamefont{Matsumoto}},
  \bibnamefont{and} \bibinfo{author}{\bibfnamefont{D.}~\bibnamefont{Villani}},
  \bibinfo{journal}{J.~Phys.~Studies} \textbf{\bibinfo{volume}{3}},
  \bibinfo{pages}{474} (\bibinfo{year}{1999}).

\bibitem[{\citenamefont{Cannon et~al.}(1991)\citenamefont{Cannon, Scalettar,
  and Fradkin}}]{Cannon:91}
\bibinfo{author}{\bibfnamefont{J.}~\bibnamefont{Cannon}},
  \bibinfo{author}{\bibfnamefont{R.}~\bibnamefont{Scalettar}},
  \bibnamefont{and} \bibinfo{author}{\bibfnamefont{E.}~\bibnamefont{Fradkin}},
  \bibinfo{journal}{Phys.~Rev.~B} \textbf{\bibinfo{volume}{44}},
  \bibinfo{pages}{5995} (\bibinfo{year}{1991}).

\bibitem[{\citenamefont{Japaridze and Kampf}(1999)}]{Japaridze:99}
\bibinfo{author}{\bibfnamefont{G.}~\bibnamefont{Japaridze}} \bibnamefont{and}
  \bibinfo{author}{\bibfnamefont{A.}~\bibnamefont{Kampf}},
  \bibinfo{journal}{Phys.~Rev.~B} \textbf{\bibinfo{volume}{59}},
  \bibinfo{pages}{12822} (\bibinfo{year}{1999}).

\bibitem[{\citenamefont{Nakamura}(2000)}]{Nakamura:00}
\bibinfo{author}{\bibfnamefont{M.}~\bibnamefont{Nakamura}},
  \bibinfo{journal}{Phys.~Rev.~B} \textbf{\bibinfo{volume}{61}},
  \bibinfo{pages}{16377} (\bibinfo{year}{2000}).

\bibitem[{\citenamefont{Tsuchiizu and Furusaki}(2002)}]{Tsuchiizu:02}
\bibinfo{author}{\bibfnamefont{M.}~\bibnamefont{Tsuchiizu}} \bibnamefont{and}
  \bibinfo{author}{\bibfnamefont{A.}~\bibnamefont{Furusaki}},
  \bibinfo{journal}{Phys.~Rev.~Lett.} \textbf{\bibinfo{volume}{88}},
  \bibinfo{pages}{056402} (\bibinfo{year}{2002}).

\bibitem[{\citenamefont{Sengupta et~al.}(2002)\citenamefont{Sengupta, Sandvik,
  and Campbell}}]{Sengupta:02}
\bibinfo{author}{\bibfnamefont{P.}~\bibnamefont{Sengupta}},
  \bibinfo{author}{\bibfnamefont{A.}~\bibnamefont{Sandvik}}, \bibnamefont{and}
  \bibinfo{author}{\bibfnamefont{D.}~\bibnamefont{Campbell}},
  \bibinfo{journal}{Phys.~Rev.~B} \textbf{\bibinfo{volume}{65}},
  \bibinfo{pages}{155113} (\bibinfo{year}{2002}).

\bibitem[{\citenamefont{Jeckelmann}(2002)}]{Jeckelmann:02}
\bibinfo{author}{\bibfnamefont{E.}~\bibnamefont{Jeckelmann}},
  \bibinfo{journal}{Phys.~Rev.~Lett.} \textbf{\bibinfo{volume}{89}},
  \bibinfo{pages}{236401} (\bibinfo{year}{2002}).

\bibitem[{\citenamefont{Nakamura}(1999)}]{Nakamura:99}
\bibinfo{author}{\bibfnamefont{M.}~\bibnamefont{Nakamura}},
  \bibinfo{journal}{J.~Phys.~Soc.~Jpn.} \textbf{\bibinfo{volume}{68}},
  \bibinfo{pages}{3123} (\bibinfo{year}{1999}).

\bibitem[{\citenamefont{Mancini}(2000)}]{Mancini:00b}
\bibinfo{author}{\bibfnamefont{F.}~\bibnamefont{Mancini}},
  \bibinfo{journal}{Europhys.~Lett.} \textbf{\bibinfo{volume}{50}},
  \bibinfo{pages}{229} (\bibinfo{year}{2000}).

\bibitem[{\citenamefont{Chattopadhyay and Gaitonde}(1997)}]{Chattopadhyay:97}
\bibinfo{author}{\bibfnamefont{B.}~\bibnamefont{Chattopadhyay}}
  \bibnamefont{and} \bibinfo{author}{\bibfnamefont{D.}~\bibnamefont{Gaitonde}},
  \bibinfo{journal}{Phys.~Rev.~B} \textbf{\bibinfo{volume}{55}},
  \bibinfo{pages}{15364} (\bibinfo{year}{1997}).

\bibitem[{\citenamefont{Avella et~al.}(1998{\natexlab{d}})\citenamefont{Avella,
  Mancini, and Villani}}]{Avella:98c}
\bibinfo{author}{\bibfnamefont{A.}~\bibnamefont{Avella}},
  \bibinfo{author}{\bibfnamefont{F.}~\bibnamefont{Mancini}}, \bibnamefont{and}
  \bibinfo{author}{\bibfnamefont{D.}~\bibnamefont{Villani}},
  \bibinfo{journal}{Sol.~Stat.~Comm.} \textbf{\bibinfo{volume}{108}},
  \bibinfo{pages}{723} (\bibinfo{year}{1998}{\natexlab{d}}).

\bibitem[{\citenamefont{Markiewicz}(1997)}]{Markiewicz:96}
\bibinfo{author}{\bibfnamefont{R.}~\bibnamefont{Markiewicz}},
  \bibinfo{journal}{J.~Phys.~Chem.~Sol.} \textbf{\bibinfo{volume}{58}},
  \bibinfo{pages}{1179} (\bibinfo{year}{1997}).

\bibitem[{\citenamefont{Torrance et~al.}(1989)\citenamefont{Torrance, Bezinge,
  Nazzal, Huang, Parkin, Keane, LaPlaca, Horn, and Held}}]{Torrance:89}
\bibinfo{author}{\bibfnamefont{J.}~\bibnamefont{Torrance}},
  \bibinfo{author}{\bibfnamefont{A.}~\bibnamefont{Bezinge}},
  \bibinfo{author}{\bibfnamefont{A.}~\bibnamefont{Nazzal}},
  \bibinfo{author}{\bibfnamefont{T.}~\bibnamefont{Huang}},
  \bibinfo{author}{\bibfnamefont{S.}~\bibnamefont{Parkin}},
  \bibinfo{author}{\bibfnamefont{D.}~\bibnamefont{Keane}},
  \bibinfo{author}{\bibfnamefont{S.}~\bibnamefont{LaPlaca}},
  \bibinfo{author}{\bibfnamefont{P.}~\bibnamefont{Horn}}, \bibnamefont{and}
  \bibinfo{author}{\bibfnamefont{G.}~\bibnamefont{Held}},
  \bibinfo{journal}{Phys.~Rev.~B} \textbf{\bibinfo{volume}{40}},
  \bibinfo{pages}{8872} (\bibinfo{year}{1989}).

\bibitem[{\citenamefont{Johnston}(1989)}]{Johnston:89}
\bibinfo{author}{\bibfnamefont{D.}~\bibnamefont{Johnston}},
  \bibinfo{journal}{Phys.~Rev.~Lett.} \textbf{\bibinfo{volume}{62}},
  \bibinfo{pages}{957} (\bibinfo{year}{1989}).

\bibitem[{\citenamefont{Loram et~al.}(1989)}]{Loram:89}
\bibinfo{author}{\bibfnamefont{J.}~\bibnamefont{Loram}} \bibnamefont{et~al.},
  \bibinfo{journal}{Physica~C} \textbf{\bibinfo{volume}{162}},
  \bibinfo{pages}{498} (\bibinfo{year}{1989}).

\bibitem[{\citenamefont{Wada et~al.}(1990)\citenamefont{Wada, Obana, Nakamura,
  and Kumagai}}]{Wada:90}
\bibinfo{author}{\bibfnamefont{N.}~\bibnamefont{Wada}},
  \bibinfo{author}{\bibfnamefont{T.}~\bibnamefont{Obana}},
  \bibinfo{author}{\bibfnamefont{Y.}~\bibnamefont{Nakamura}}, \bibnamefont{and}
  \bibinfo{author}{\bibfnamefont{K.}~\bibnamefont{Kumagai}},
  \bibinfo{journal}{Physica~B} \textbf{\bibinfo{volume}{165-166}},
  \bibinfo{pages}{1341} (\bibinfo{year}{1990}).

\bibitem[{\citenamefont{{J.W. Loram} et~al.}(1993)}]{Loram:93}
\bibinfo{author}{\bibnamefont{{J.W. Loram}}} \bibnamefont{et~al.},
  \bibinfo{journal}{Phys.~Rev.~Lett.} \textbf{\bibinfo{volume}{71}},
  \bibinfo{pages}{1740} (\bibinfo{year}{1993}).

\bibitem[{\citenamefont{Loram et~al.}(1994)}]{Loram:94}
\bibinfo{author}{\bibfnamefont{J.}~\bibnamefont{Loram}} \bibnamefont{et~al.},
  \bibinfo{journal}{Physica~C} \textbf{\bibinfo{volume}{235-240}},
  \bibinfo{pages}{134} (\bibinfo{year}{1994}).

\end{thebibliography}

\end{document}